\documentclass[preprint,floats,eqsecnum,aps,tightenlines,showpacs]{revtex4}
\usepackage{epsfig}

\newcommand{\be}{\begin{equation}}
\newcommand{\ee}{\end{equation}}
\newcommand{\bea}{\begin{eqnarray}}
\newcommand{\eea}{\end{eqnarray}}
 
\newcommand{\nn}{\nonumber}

\begin{document}

\preprint{ \parbox{1.5in}{\leftline{hep-th/0111433}}}

\title{\bf Study of relativistic bound states for scalar theories 
in Bethe-Salpeter and Dyson-Schwinger formalism}

\author{Vladim\'{\i}r \v{S}auli,  J.\ Adam, Jr.}  

\affiliation{Department of Theoretical Physics, 
Nuclear Physics Institute, \v{R}e\v{z} near Prague, CZ-25068,
Czech Republic}

\begin{abstract}

The Bethe-Salpeter equation for Wick-Cutkosky like models is solved in  
dressed ladder approximation. The bare vertex truncation of the 
Dyson-Schwinger  equations for propagators is combined with the dressed 
ladder Bethe-Salpeter equation for the scalar S-wave bound state 
amplitudes. With the help of spectral representation the results are 
obtained directly in Minkowski space. We give a new  analytic formula 
for the resulting equation simplifying the numerical treatment. The 
bare ladder approximation of Bethe-Salpeter equation is compared with 
the one with dressed ladder. The elastic electromagnetic form factors 
is calculated within the relativistic impulse approximation. 
\end{abstract}

\maketitle
%

\section{Introduction}
 
In the Quantum Field Theory the two body bound state is described by 
the three-point bound state vertex function or, equivalently, by 
Bethe-Salpeter (BS) amplitudes, both of  them are solutions of the 
corresponding (see Fig.~\ref{figBSE}) covariant four-dimensional 
Bethe-Salpeter equations (BSE)\cite{BETHE}.  Up to now most of the 
studies  were restricted  to the case when irreducible interaction 
kernel is  approximated by (sum of) single particle exchanges. In this 
so called ladder approximation the  scattering matrix is given  by the 
sum of the generated ladders.   It is  known, that such approximation 
is not sufficient when more realistic models  are considered  
\cite{ROBERTS}, \cite{SMEKAL}, \cite{SCHMIDT}. To move beyond this 
approximation  one is  in practice  confined to the use of  some 
phenomenological Ansatze.  (In hadronic physic, these Ansatze are very 
often made already at the level of two point correlators. For modeling 
of the gluon propagator in the context of BSE and  Dyson-Schwinger 
equations (SDE), see for instance \cite{MARIS}).  

Here we are considering some simple scalar models. An extensive review 
of BS studies in scalar theories (with at most cubic and non-derivative 
interaction) can be found in Ref.\ \cite{Nakan}, \cite{SETO}.  The 
various improvements of simple ladder kernel have been considered, in 
particular, including the self-energy effects \cite{KUSAK3}, 
\cite{AHLIG} or contributions from crossed box diagrams \cite{THEUSL}. 
The study of the influence on the bound state spectrum  following from 
the infinite resummation of certain ladder and crossed-box diagrams can 
be found in the paper \cite{TJON2}. Furthermore, there is a number of 
interesting papers  on the solution  of  Wick-Cutkosky models (with 
zero mass of the exchanged particle). These solutions employ various 
effective techniques like the point form of relativistic quantum 
mechanics \cite{DEPLANK}, variational calculations \cite{DAREWYCH} or 
the light front dynamics \cite{MILLER}. 

The standard approach to determine the spectrum and  the  BS vertex  
makes use of partial wave decomposition which reduces four-dimensional 
integral equation into the two-dimensional one. The alternative  more 
recently exploited treatment is based on the $O(4)$ hyperspherical  
expansion  \cite{TJON}. In this approach the BSE is transformed into an 
infinite set of one-dimensional  integral equations. Notable advantage 
of this approach is a good numerical convergence  and easy 
identification of excited bound states spectra. 

Very often, the ladder BSE's are solved with the help of  the so-called 
Wick rotation \cite{WICK}. However, the backward analytical 
continuation is quite difficult even for the  ladder approximation, 
while for more complicated cases its proper implementation  is unclear 
or at least highly non-trivial. 

In this work we follow the method of solving the  BSE  directly in 
Minkowski space \cite{KUSAK3}, \cite{KUSAK1}, \cite{FULTON},  in which 
the problems associated with Wick rotation do not arise.  The method is 
based on utilization of  generalized spectral representation for 
n-point Green functions in quantum field theory \cite{NAKANATO}. In 
this treatment the BSE written in momentum space is converted into a 
real integral equation for a real weight function with number of 
independent variables dependent on details of the model.  We extend the 
earlier work \cite{KUSAK3} first to the case with unequal masses of 
constituents. This then allows us to treat the ladder BSE in which all 
propagators (of constituents and of the exchanged  particle) are fully 
dressed. This is achieved by the implementation of  Lehmann 
representation of the propagator: 
\bea        
G(p^2)&=&\int\limits d\omega\, \tilde{\sigma}(\omega)D(p;\omega)\, ,
 \nn \\
D(p;\omega)=\frac{1}{p^2-\omega+ i\epsilon} \quad &;& \quad
\tilde{\sigma}(\omega)= R\, \delta(\omega-m^2_{pole})+\sigma(\omega)\, .
\label{spektrato}
\eea
where $\sigma(\omega)$, which is a smooth function, nonzero above
a threshold, is determined by the Dyson-Schwinger equations 
(DSE). In the pole term we can take  $R=1$, a choice corresponding to 
the  conventional on-shell renormalization scheme ($G$ has a unit 
residuum when momentum approaches  a simple pole at physical mass). 
 
To account for the effect of self-energy we transform the momentum BSE 
to the form suitable for complementary solution together with the  
appropriate DSE for propagators. Note here, that the perturbative one 
loop  contribution has been already considered in   \cite{KUSAK3} and 
certain  Euclidian  version of this problem has also been  
 investigated \cite{AHLIG}. In qualitative agreement with 
\cite{AHLIG} we have found that the critical value of coupling  gives 
the domain of applicability of  BSE (at least in its ladder 
approximation). The couplings below critical one allow only solutions 
for  relatively weakly bound states. It is even more interesting that  
the effect of the propagator dressing on bound state spectra is rather 
small. In comparison with the bare ladder approximation the same 
binding energy is then achieved with the coupling smaller by about 
several per cents, even for values of the coupling close to the 
critical one. 

Clearly, when we take some or even all particle propagators dressed, the 
number of spectral integrations  increases.  Note here, that up to the 
rather exotic case of massless Wick-Cutkosky model the appropriate 
solution is not known analytically but must be found numerically. 
Mainly due to this reason  we re-formulate the equation  obtained by 
Kusaka et al \cite{KUSAK3} and we offer the solution where the  
appropriate integral kernel is  free of any additional numerical 
integration (see Appendix of Ref.~\cite{KUSAK3} for the original 
solution). The elimination of this numerical integration then not only 
improves numerical accuracy but also reasonably decreases the CPU time. 

To see explicitly the effect of radiative corrections  we  compare the 
derssed BSE results with its bare ladder approximation. We set the 
parameters of our model to  that used in Refs.~\cite{TJON} and 
\cite{KUSAK1}, to compare the bare ladder solutions  to those  obtained 
before in \cite{KUSAK3}, \cite{TJON}, \cite{KUSAK1}. 

Having solved equations for spectral functions, one can determine 
the BS amplitudes in an arbitrary reference frame. This makes this 
technique suitable for calculations of response to an external fields. 
In section IV. we briefly introduce the formulas defining the elastic 
charge form factor $G(Q^2)$ in relativistic impulse approximation 
(RIA). Although the elastic form factor represents a simple dynamical 
observable, its Minkowskian calculation represents nontrivial task. For 
this purpose we consider the (massive)  Wick-Cutkosky model given by  
Lagrangian  gauged as follows: 
\bea \label{gaugedwcm}
{\cal L}&=& (D^{\mu}\phi_1)^+ D_{\mu}\phi_1
+\frac{1}{2}\partial_{\mu}\phi_2\partial^{\mu}\phi_2
+\frac{1}{2}\partial_{\mu}\phi_3\partial^{\mu}\phi_3-
\frac{1}{4}F_{\mu\nu}F^{\mu\nu} -V(\phi_i) \, ,
\nn \\
V(\phi_i)&=& \left(m_1^2+g\phi_3\right)\phi_1^+\phi_1 +
\left( \frac{m_2^2}{2}+
{g \over 2}\, \phi_3\right)\phi^2_2+\frac{1}{2}m_3^2\phi_3^2 \, ,  
\eea
where covariant derivative is $D_{\mu}=\partial_{\mu}-ieA_{\mu}$. In 
our form factor calculations the effects of scalar dressing was not 
taken into account, since it would significantly increase computational 
complexity of the problem. Furthermore, it is assumed that $e<<g$ which 
implies that the interaction of the charged particle field $\phi_1$ 
with the electromagnetic field can be treated perturbatively. As in 
Ref.~\cite{AHLIG}, we have chosen the same coupling constant
for interaction of the field $\phi_3$ with the fields $\phi_1$ and
$\phi_2$. The form factors were calculated for several sets of masses
of constituent and exchanged particles.

\section{Dressed Ladder Bethe-Salpeter Equation}

The BS amplitude for bound state $(\phi_1,\phi_2)$ in momentum space is 
defined through the Fourier transform of 
\be
\langle 0|T\phi_1(x_1)\phi_2(x_2)|P\rangle
    = e^{-iP\cdot X}\langle 0|T\phi_1(\eta_2 x)\phi_2(-\eta_1 x)|P\rangle
    = e^{-iP\cdot X}
    \int{d^4p\over (2\pi)^4} e^{-ip\cdot x} \Phi(p,P)\ ,
\ee
where $X\equiv \eta_1x_1+\eta_2x_2$ and $x\equiv x_1-x_2$, so that 
$x_1=X+\eta_2x$, $x_2=X-\eta_1x$. Here $p_{1,2}$ are the  four-momenta 
of particles corresponding to the fields  $\phi_{1,2}$ that constitute 
the bound state  $(\phi_1,\phi_2)$.  The total and relative momenta are 
then given as $P= p_1 +p_2 $ and $ p=(\eta_2p_1-\eta_1p_2)$, 
respectively, and $P^2=M^2$, where $M$ is mass of the bound state. 
Finally, $P\cdot X + p\cdot x = p_1\cdot x_1 + p_2\cdot x_2$. From now 
on we will put $\eta_1= \eta_2= 1/2$, which corresponds to the usual 
separation of center of mass motion for equal mass case, but can be 
also employed for unequal masses (although $X$ is then not the 
coordinate of the center of mass). 
 
Introducing the BS vertex function $\Gamma=iG_1^{-1}G_2^{-1} \Phi $, 
the homogeneous  BSE for S-wave bound state reads 
\be  \label{bse}
\Gamma(p,P) = i\int{d^4k\over(2\pi)^4}\,
V(p,k;P) G_1(k+P/2)G_2(-k+P/2) \Gamma(k,P)\, . 
\ee
The bound states appear as  poles of the scattering matrix. The 
normalization condition for the BS vertex function follows from the 
requirement that the  pole appropriate to a given bound state is  a 
simple one: 
\bea      
\label{normalization}
2iP^\mu\ &=& \int \frac{d^4p}{(2\pi)^4}\int \frac{d^4k}{(2\pi)^4}
    \bar\Gamma(k,P) \left[ (2\pi)^4\delta^4(p-k)
\left( \frac{\partial}{\partial P_\mu}G_1(p_1)G_2(p_2)\right) 
\right. \nonumber\\
&& \left. + i\, G_1(p'_1) G_2(p'_2)\, 
\left( \frac{\partial}{\partial P_\mu}V(p,k;P)\right)\,
G_1(p_1) G_2(p_2)\,  \right]\, \Gamma(p,P) \, ,
\eea
where $\bar\Gamma(p,P)$ is the conjugate of $\Gamma(p,P)$ and $p_{1,2}= 
\pm p + P/2, p'_{1,2}= \pm k + P/2$ (for the details see e.g. 
Ref.~\cite{Nakan}). 

In this work we do not solve the  BSE with most general irreducible 
scattering kernel $V$ (the most general structure of the kernel $V$ 
written in terms of its Perturbation Theory Integral Representation 
(PTIR)  can be found in Ref. \cite{KUSAK3} or  
\cite{NAKANATO}). Here we restrict ourselves to the case of dressed 
ladder approximation with $\phi_3$-exchange, in which 
$V(p,k;P)= g^2 G_3(t)$, where $t$ denotes the usual Mandelstam variable  
$t=(p-k)^2$.  Note that for the bound state of particles 
$(\phi_1,\phi_2)$ only the $t$-channel interaction above is effective, 
whereas for the bound state of  $(\phi_i,\phi_i), i=1,2$ one has to 
consider also possible $u$ and $s$  channel diagrams. In the present 
work we study the case $(\phi_1,\phi_2)$ due to  its simplicity and 
leave the other cases for discussion elsewhere. Let us also recall that 
in the (dressed) ladder approximation, defined by the exchange of 
chargeless scalar $\phi_3$, the photon coupling to $\phi_1$ alone (one 
particle current, or in other words RIA) is by itself gauge invariant, 
when taken between the corresponding solutions of BSE. And the 
normalization condition (\ref{normalization}) in this approximation 
reduces to the condition $G(Q^2)=1$. We will solve the BSE for massive
constituents ($m_{1,2} > 0$) and $m_3 \geq 0$, the invariant mass of the
bound state satisfies $0\leq P^2 < (m_1+ m_2)^2$. 

Taking the dressed kernel and  the full propagators of constituent 
particles  into account, the rhs of BSE (\ref{bse}) can be written as 
\bea  \label{bse2} 
i g^2 \prod_{i=1}^3\int d\alpha_i\, 
\tilde{\sigma}(\alpha_i)\int{d^4k\over(2\pi)^4} 
D(k+P/2;\alpha_1) D(-k+P/2;\alpha_2) D(k-p;\alpha_3) \Gamma(k,P)\, . 
\eea 
The interesting unequal-mass ladder case of Ref.~\cite{FUKUI} is also 
described by Eq.(\ref{bse2}), although $\eta_1= \eta_2= 1/2$ and $p$ 
and $k$ are not relative momenta. Since the dependence on momenta in
(\ref{bse2}) is explicit, one might always re-scale $\Gamma(p,P)$ to
proper relative momentum.    

The  integral representation for the BS vertex function may be written 
as \cite{NAKANATO} 
\be  \label{nula}
\Gamma(p,P)= \int\limits_{-1}^1 dz \, 
\int\limits_{\alpha_{min}(z)}^{\infty}d\alpha\, \,
  \frac{\rho^{[n]}(\alpha,z)}{[\alpha-(p^2+zp\cdot P+P^2/4)-i\epsilon]^n} \, \, .
\ee
The positive integer $n$ represents a free parameter without clear 
physical meaning. One can take advantage of this freedom of choice to 
pick up $n$ so that the numerical solutions of integral equations for 
spectral functions are made more stable. The spectral functions 
$\rho^{[n]}(\alpha,z)$ for different $n$ can be related by integration 
over $\alpha$ by parts. Kusaka et al \cite{KUSAK3} choose $n=2$ for 
their numerical solution of the BSE, we adopt the same value in this 
paper. 

The bare (symmetric) Wick-Cutkosky model  corresponds  to the choice: 
$\alpha_1=\alpha_2=m^2$, the exchanged boson is massless ($\alpha_3=0$) 
and no radiative corrections are considered. This model is particularly 
interesting because it is the only example of the nontrivial BSE which 
is solvable exactly  \cite{WICK}. For this model, there is no freedom 
in choice of $n$ and (for the S-wave bound state) the expression 
(\ref{nula}) reduces to the one dimensional PTIR: 
\be  \label{wcm}
\Gamma(p,P)=\int\limits_{-1}^1 dz \,
\frac{\rho(z)}{m^2-(p^2+zp\cdot P+P^2/4)- i\epsilon} \, \, .
\ee  

Using a  technique similar to the one used in ref.~\cite{KUSAK3}, the 
BSE can be converted to the following real integral equation for the 
real spectral function: 
\be   \label{majn}
\rho^{[n]}(\alpha',z')=\lambda \int\limits_{-1}^{1}dz
\int\limits_{\alpha_{min}(z)}^{\infty} 
d\alpha\, \, V^{[n]}(\alpha',z';\alpha,z)\, \rho^{[n]}(\alpha,z)  \, ,
\ee
where we denoted $\lambda=g^2/(4\pi)^2$. The  derivation is presented 
in the Appendix A, where the explicit expressions for particular 
choices $n=1,2$ are given. The central results of this work are  
expressions obtained for $V^{[n]}$, which are simpler that the ones 
presented in Ref.~\cite{KUSAK3}. No additional integration is required 
which decreases the computer time necessary for numerical calculation. 
Besides, our formulas hold also for unequal masses of the constituents. 
The extension (together with above mentioned simplification) to the 
case of more complicated scattering kernel is not so straightforward, 
but we believe that it is possible. Note also, that due to the property 
of solid harmonic with respect to the integration over the momentum the 
presented procedure can  easily be generalized for the bound state with 
non-zero spin (here, the total orbital momentum)\cite{KUSAK3}. 
   
A bound state with equal composite masses is described by the vertex 
function $\Gamma$ which is symmetric under the transformation $P\cdot p 
\rightarrow -P\cdot p$. In terms of the  weight function this symmetry 
reads $\rho(\alpha,z)=\rho(\alpha,-z)$.  However, there are solutions 
that do not respect this symmetry even in the case of equal masses. 
These are usually called ghost solutions and the appropriate  
amplitudes have a negative norm. Such  solutions are often considered 
to be  nonphysical and it is supposed that they point at inner 
inconsistency in description of relativistic bound states within the BS 
formalism, at least in the ladder approximation. At this place, it is 
important to mention that the Lagrangian  (\ref{gaugedwcm}) describes 
the models that are a subset of theories with potentials unbounded from 
below and in very strict sense they are discarded due to the vacuum 
instability. On the other hand one can assume, at least for 
sufficiently small couplings the existence of local minima of the 
potentials is sufficient to  support of the existence of ground state 
of the theory. While in the large coupling regime, say for $g/m>>1$ no 
reasonable physics can be learned from the perturbation theory and/or 
from equation like ladder BS by itself. To conclude, we note  that such 
ideas are supported by at least two facts. The ghost BS solutions do 
appear only for a large value of $\lambda$. Furthermore, from the 
Dyson-Schwinger study we know that the scalar theory studied here makes 
sense only up to the certain critical coupling, see e.g. \cite{AHLIG}, 
\cite{SUPER}. 

The  important question arises, what is the validity of the full theory 
when renormalization is properly taken into account. Although the  
quantitative  answer lies beyond the approximations used in this 
article and requires more careful investigation, we make a simple 
attempt to find the domain in which self-consistent solutions of the 
BSE and Dyson-Schwinger equations (within the framework of reasonable 
approximations) exist. Furthermore, we inquire an influence of scalar 
propagator dressing on the solution of the BSE for the bound states. 

\section{Dressing propagators by the Dyson-Schwinger equations } 

The solution of the DSE for the scalar models with the help of spectral 
decomposition will be discussed in detail in our forthcoming paper 
\cite{SUPER}. Here we give only brief presentation of the DSE in bare 
vertex approximation, their renormalization, re-arrangement in terms of 
the spectral function and some properties of solutions, important for 
our further discussion of the BSE. In this section we assume $m_3 < 
m_1+ m_2$, so that the propagator of the exchanged particle $\Phi_3$ 
has an isolated physical pole. 

By dressing of the scalar propagators in our study of BSE we mean only 
the dressing due to the ``strong'' interaction between scalars, 
the coupling to the electromagnetic field is neglected. Let us now write the
strong interaction part of our Lagrangian (\ref{gaugedwcm}) in terms
of bare, unrenormalized quantities (fields and coupling constants), 
labeled by subscript "0": 
\bea \label{vcml}
{\cal L}_{strong}= - g_{01}\, \phi_{01}^+ \phi_{01} \phi_{03}
 - { g_{02} \over 2}\, \phi_{02}^2 \phi_{03} \,  .
\eea
In the previous section we have chosen the strength of of both couplings
to be the same. Here, we distinguish the bare couplings, anticipating that
they are renormalized by different amounts (see below).

The kinetic terms are parametrized by the unrenormalized masses 
$m_{0i}$. These masses undergo the {\em infinite} mass renormalization 
\be
 m_{0i}^2= m_i^2- \delta m^2_i \, , \quad i=1,2,3 \, .
\label{rmasses} 
\ee
To re-scale the residuum of the full propagators to unity, we will 
complement the infinite mass renormalization by the {\em finite} (since 
the model is superrenormalizable) renormalization of the fields and 
coupling constants 
\be
\label{konstanty}
\phi_{0i}= \sqrt{Z_i}\, \phi_{i} \, , \quad i= 1,2,3 \, , \quad \quad   
g_{0i}=\frac{1}{Z_i \sqrt{Z_3} }\, g_i,  \quad i=1,2 \, .
\ee 
That is, we will employ below the on shell renormalization scheme in 
which the propagators have unit residua when momentum approaches  its 
mass shell value $p^2\rightarrow m^2$. 
 
In this paper we consider the Dyson-Schwinger equations in the simplest 
approximation in which the proper vertices are replaced by the bare 
ones $\Gamma_{oi}=g_{oi}$. Then, the DSE in their unrenormalized form 
read  
\bea  \label{pipin}
  G_{0i}^{-1}\left( p\right) &=& p^2-m_{0i}^{2}-\Pi_{0i}(p^2) \, , 
  \quad \quad i=1,2,3 \nonumber \\
\Pi_{0i}(p^2)&=& i\, g_{0i}^{2}\int \frac{d^{4}q}{\left( 2\pi \right) ^{4}}
\,G_{03 }\left( p-q\right) G_{0i}\left( q\right) \, , 
\quad \quad  \quad i=1,2 \nonumber \\
\Pi_{03}(p^2)&=& i\int \frac{d^{4}q}{\left( 2\pi \right) ^{4}}\,
  \sum_{i=1,2}g_{0i}^{2}
  G_{0i}\left( p-q\right) G_{0i}\left( q\right) \, ,
\eea
where $G_0(p)$ is the Fourier transform of  the full unrenormalized 
propagator $ G_{0i}(x-y)= $ $<0|T \phi_{0i}(x)\phi_{0i}(y)|>$ and 
$\Pi_{0i}$ is the corresponding self-energy. 

Under the field strength renormalization the propagators scale like 
$G_{0i}=Z_i G_i$. Multiplying the equations for $G_{0i}^{-1} $ in 
(\ref{pipin}), defining $\Pi_i= Z_i \Pi_{0i}$ and making use of 
(\ref{konstanty}), one gets the re-scaled DSE 
\bea
 G_{i}^{-1}(p^2)&=& Z_i(p^2-m_{0i}^{2})-\Pi_i(p^2) \, ,
\nn \\
\Pi_i(p^2)&=& i\, g_{i}^{2}\int \frac{d^{4}q}{\left( 2\pi \right) ^{4}}
\,G_{3 }\left( p-q\right) G_{i}\left( q\right) \, , 
\quad \quad  \quad i=1,2 \nonumber \\
\Pi_3(p^2)&=& ig^{2}\int \frac{d^{4}q}{\left( 2\pi \right) ^{4}}\,
  \sum_{i=1,2}
  G_{i}\left( p-q\right) G_{i}\left( q\right) \, .
\label{rescale}  
\eea
The renormalization of proper self-energies proceeds by double subtraction:
\be  \label{of} 
\Pi_{iR}(p^2)=\Pi_i(p^2)- \Pi_i(m^2)-
 (p^2-m_i^2)\, \frac{d\Pi_i(p^2)}{dp^2}|_{p^2=m_i^2}\, \, . 
\ee 
Identifying the appropriate renormalization constants 
(\ref{rmasses},\ref{konstanty}) 
\bea \label{podminky}
\delta m^2_i=\Pi_i(m_i^2)/Z_i \, , \quad \quad
Z_i=1+\frac{d\Pi_i(p^2)}{dp^2}|_{p^2=m_i^2} \, ,
\eea
we can immediately  write the full propagator in terms of finite
physical quantities
\be  \label{jednadva}
  G_{i}^{-1}\left( p\right) = p^2-m_i^{2}-\Pi_{iR}(p^2) \, ,  \quad i=1,2,3 \, .
\ee 
The DSE for the renormalized self-energies are given by (\ref{rescale}) 
and subtraction (\ref{of}). 

For the purpose of our BS calculation, we now fix the renormalized 
couplings and masses as follows 
\begin{equation}
 g_{1}=g_{2}\equiv g ;\quad    m_1=m_2\equiv m ; \quad  m_3=\frac{m}{2}
\label{couplings}
\end{equation}
where $g$ is coupling constant from (\ref{gaugedwcm}). That is, we will 
compare the solutions of the BSE for the bare and dressed ladder kernel 
taken for the same numerical value of unrenormalized and renormalized 
coupling constant, respectively. The masses   are fixed to allow 
comparison with some of the results of Refs.~\cite{KUSAK1,KUSAK3,TJON}. 
  
Now, it  is straightforward task to evaluate the spectral 
representation of the renormalized self-energy.  
Lehmann representation (with unit residuum) for $G_i$ reads: 
\be
 G_i(p^2) = \int_{0}^\infty 
 ds \frac{\tilde{\sigma}(s)}{p^2- s+ i \epsilon}\, , \quad \quad
\tilde{\sigma}(s)=\delta(m_i^2-s)+\sigma(s) \, , 
\label{introg}
\ee 
Notice that functions $\sigma_i, i= 1,2$, have the thresholds at 
$m_{i,th}=(m_i+ m_3)^2= 2.25m^2$, whereas the function  $\sigma_{3}$ 
has the threshold at $m_{3,th}= (m_1+m_2)^2= 4m^2$. Analogously, for the for 
self-energies: 
\begin{equation} \label{intro}
\Pi_{iR}(p^2)=  \int_{m_{i,th}}^\infty  d\alpha\,
\frac{\rho_{\pi i}(\alpha)}{p^{2}-\alpha+i\epsilon}
\frac{(p^2-m^2)^2}{(\alpha-m^2)^2} \, .
\end{equation}
The spectral representation for $\Pi_R$ explicitly satisfies $\Pi_R(m^2)= 
\Pi_R^{\prime} (m^2)=0$ following from (\ref{of}). Rewriting now the 
relation between $G$ and $\Pi$ in the form $G= D+ D\Pi G$ ($D$ being 
the free propagator with the physical mass) and taking its imaginary part, 
we arrive to the first relation between the spectral functions $\sigma$ 
and $\rho$: 
\begin{equation}    \label{SYMB3}
\sigma_{i}(\omega)=\frac{\rho_{\pi_i}(\omega)}{(\omega - m_i^2)^2}+
(\omega - m_i^2)\, P\int {d\alpha \over \omega-\alpha}
\left[ \frac{\sigma_i(\omega) \rho_{\pi_i}(\alpha)}{(\alpha - m_i^2)^2}+
\frac{\sigma_i(\alpha) \rho_{\pi_i}(\omega)}{(\omega - m_i^2)^2} \right] 
\, , \quad i=1,2,3 \, ,
\end{equation}
where  $P\int$ stands for principal value integration. All the 
functions in (\ref{SYMB3}) are positive and regular above the 
perturbative thresholds and identically equal to zero elsewhere. 

Substituting the spectral representations (\ref{introg}) into the DSEs 
(\ref{rescale}), making the subtraction as in (\ref{of}) and comparing 
to the lhs in the form of (\ref{intro}),  one gets after lengthy 
algebra: 
\bea \label{vysl}
\rho_{\pi_i}(\omega)&=& \lambda\,
\int d\alpha\, d\beta\, 
B(\alpha,\beta;\omega)\tilde{\sigma}_3(\alpha)
\tilde{\sigma}_i(\beta) \, ,\quad \quad i=1,2 \, ,
\nn \\
\rho_{\pi_3}(\omega)&=& \lambda\, 
\sum_{i=1,2}\int d\alpha\, d\beta\,
B(\alpha,\beta;\omega)\tilde{\sigma}_i(\alpha) \tilde{\sigma}_i(\beta) 
\, ,
\eea
where  $\lambda=g^2/(4\pi)^2$ and the function $B(\alpha,\beta;\omega)$ 
is related to the K\"allen function $\lambda$ as follows: 
\begin{eqnarray}
B(\alpha,\beta,\omega)&=&\frac{\sqrt{\lambda(\alpha,\beta,\omega)}}{\omega}\,
\Theta\left( \omega-(\sqrt{\alpha}+\sqrt{\beta})^2 \right) \, ,
\nonumber \\
\lambda(\alpha,\beta,\omega)&=& \alpha^2 + \beta^2 + \omega^2-
 2 \alpha \beta - 2 \alpha \omega -2 \beta \omega  , .
\end{eqnarray}
Before the numerical treatment the explicit integration- separating 
the $\delta$-function parts of Lehmann weights  $\tilde{\sigma}$-
has to be performed. 

Equations (\ref{SYMB3},\ref{vysl})  constitute the closed system of 
integral equations for spectral functions which can be solved 
numerically by iterations without any additional approximation. So 
obtained dressed propagators have been used when solving the 
Bethe-Salpeter equation. The results are discussed in the section V. Before 
leaving this section we review some important features of our solutions 
of DSE. 

The behavior of the imaginary parts of propagators-- the Lehmann 
functions $\sigma_i(\alpha)$-- for fields $\Phi_{1,2,3}$ is shown at 
Fig.~\ref{figDSEsig}. 
     
The renormalization constant $Z_i$ are calculated from the relation 
\be
Z_i=1-\int d\alpha\, \frac{\rho_i(\alpha)}{(\alpha-m_i^2)^2} \, .
\ee
From the Fig.\ref{figZs} we can see that   the field renormalization 
constant $Z_{1,2}$ changes sign from positive to negative at some 
critical point $\tilde{\lambda}_{crit}=g_{crit}^2/(4\pi m)^2 \simeq 
1.5 \pm 0.1$, where the error reflects the difficulty of making the 
numerical estimate of the value for which the solution cannot be found 
and the dimensionless coupling is defined as $\tilde{\lambda}= 
\lambda/m^2$ . We did not find any numerical solutions of DSE for 
couplings larger than $\tilde{\lambda}_{crit}$. It is reasonable to 
suppose that the quanta associated with the fields $\phi_{1,2}$ do not 
describe physical particles when $\tilde{\lambda}> 
\tilde{\lambda}_{crit}$.

\section{Elastic Electromagnetic Form Factor}

The electromagnetic form factors parameterize the response of bound 
systems to external electromagnetic field. The calculation of these 
observables within the BS framework proceeds along the Mandelstam's 
formalism \cite{MANDEL}. For the elastic scattering on the S-wave bound 
state, ($P_{i}^2=P_{f}^2=M^2$) the current conservation implies  the 
parameterization of the current matrix element $G^{\mu}$ in terms of 
the single real form factor $G(Q^2)$ 
\be  \label{formf}
G^{\mu}(P_f,P_i)=G(Q^2)(P_i+P_f)^{\mu} \, .
\ee
The elastic electromagnetic form factor $G(Q^2)$ depends only on the 
square of photon incoming momentum $q$ and we use the usual  SLAC 
convention $Q^2=-q^2$, so that  $Q^2$ is positive for  elastic 
kinematics. 

The matrix element of the current in relativistic impulse approximation 
(RIA) is diagrammatically depicted in Fig.~\ref{figGmu}. In this paper
we are not taking into account the dressing of the scalar propagators when
calculating the charge form factor. Then, the matrix element is  given in 
terms of the BS vertex functions as
\bea  
G^{\mu}(P+q,P)&=& i\, \int\frac{d^4k}{(2\pi)^4}\,  
\bar\Gamma(k+\frac{q}{2},P+q) \nonumber\\ 
&& \left[ D(p_f;m_1^2) j_1^{\mu}(p_f,p_i) D(p_i;m_1^2) \,
D(-k+P/2;m_2^2) \right] \Gamma(k,P) \, ,
\label{chargeff}
\eea
where we denote $P=P_i$ and $j_1^{\mu}$ represents one-body current for 
particle $\phi_1$, which for the bare particle reads  
$j_1^{\mu}(p_f,p_i)= p_f^{\mu}+p_i^{\mu}$, where $p_i,p_f$ is initial 
and final momentum of charged particle inside the loop in 
Fig.~\ref{figGmu}, i.e., $p_i= k+ P/2, p_f= q+ k+ P/2$. 

We have already mentioned in the previous section, that if the vertex 
functions $\Gamma$ are solution of the BSE with a kernel corresponding 
to exchange of single chargeless particle, the RIA defined above is by 
itself gauge invariant and the normalization condition for the BS 
amplitudes is equivalent to the normalization $G(0)=1$. 

The main result of this paper, as far as charge form factor is 
concerned, is re-writing of the rhs of Eq.~(\ref{chargeff}) directly in 
terms of the spectral weights of the bound state vertex function. It 
allows the evaluation of the form factor by calculating the integral of 
nonsingular expression, without having to reconstruct the vertex 
functions $\Gamma(p,P)$ from their spectral representation. The 
derivation of this integral involves some lengthy algebra and is 
relegated to Appendix B. 

\section{Numerical Results}

\subsection{Bare Ladder BSE}

We have solved the bare ladder BSE for symmetric ($m_1=m_2=m$) scalar 
theory with bare ladder kernel
\be
  V(p,k,P)= V(p-k)= \frac{g^2}{(p-k)^2-m^3_2} \, ,
\ee
and bare constituent propagators $G_i(p_i)= D_i(p_i,m_i)$ by iterations 
of the integral equation for spectral functions. The standard procedure 
was followed: after fixing the bound state mass ($P^2$) we looked for 
the solution by iterating  spectral function for fixed dimensionless 
"coupling strength" $\tilde{\lambda} \equiv g^2/((4\pi\,)^2 m^2)$. If 
the iterations failed-- measure being both the difference of the rhs 
and lhs of the integral equation and deviation of the auxiliary 
normalization integral from pre-defined value-- we were changing 
$\tilde{\lambda}$ (halving intervals of successive guesses) till the 
solution was found.   
 
In case of Wick-Cutkosky model the one dimensional integral equation 
(\ref{wcmsolve}) was solved. Although the solution of this 
one-dimensional integral equation could be found by the inversion of 
its discretized form, we have tested the iteration procedure (used 
later also for massive exchange). The equation was discretized by 
numerical Gauss integration, it appear that it is sufficient to take 40 
Gauss point (though the number cited in Table 1 are obtained with 98 
points). It is known \cite{Nakan} that  $P^2=0$ corresponds to 
$\tilde{\lambda}=2$ from which our result slightly deviates in the 
fifth digit. We have also reproduced (up to four published digits) all 
results for Wick-Cutkosky model from \cite{TJON}. In the 
Fig.\ref{figDSErho} the weight functions are plotted against spectral 
variable $z$ for several fractions of binding $\eta=\sqrt{P^2}/2m$ .

For the massive scalar exchange the two-dimensional integral 
equations (\ref{baren1solve},\ref{baren2solve}) were solved.
We have found (in agreement with \cite{KUSAK3}) that numerical errors 
are about one order of magnitude bigger for $n=1$, hence $n=2$ is
preferable and only the results with this choice are discussed below.
 
For numerical solution we  discretize integration variables  $\alpha$ 
and $z$ using Gauss-Legendre quadratures (with tangent mapping from 
$<-1,+1> \rightarrow <\alpha_{min},\infty >$ for $\alpha$ ). The  
equation (\ref{baren2solve}) is solved on the net of $N=N_z*N_{\alpha}$ 
points which are spread on the rectangle $(-1,+1)*(\alpha_{min}, \infty )$. 
The value $\alpha_{min}$ is  given by the support of the spectral 
function (see the appendix A). We have not optimized the grid during 
the iteration procedure as  it was done in the study \cite{KUSAK3}. 
Instead, we have solved the equation for several different numbers of 
grid points while keeping fixed the ratio of $ N_{\alpha}/N_z$ and then 
extrapolated the results to the ``ideal'' case with $ N_{\alpha}= N_z= 
\infty$. Examples of numerical convergence for some cases of bound 
states are presented in the Table 2. In the Table 1 we compare our 
results for $m_3=m/2$ with those of ref. \cite{KUSAK3}. 
 
Below we show the dependence of charge form factor
on the parameters of the model: on the range of interaction 
characterized by the inverse mass of exchanged meson  $m_3$ 
and on the strength of forces which bind the  particles together.
To calculate the form factors we first have to solve the BSE for
chosen sets of parameters. We vary the parameters as follows:
\begin{enumerate}         
\item First we solve the BSE for several
 bound state masses $P^2$ keeping the  ratio $m_3/m$   
 fixed  
\item 
Then we vary the mass of the exchanged meson, keeping the masses of all 
studied bound states fixed (our choice is $\eta=\sqrt{P^2}/(2m)=0.5$) 
and determining the corresponding coupling strengths  
$\tilde{\lambda}$. 
\end{enumerate}
Where the independent numbers were available \cite{TJON}, they agree 
with our results (see Table 3). If the mass of exchanged particle 
becomes small (but nonzero) the convergence of our numerical procedure 
becomes somewhat poorer and more sensitive to the initial guess. 
For illustration the  weight functions $\tilde{\rho}^{[2]}$  is
plotted in figs.(\ref{figBSEbarelad1},\ref{figBSEbarelad2}) for two
different values of exchanged mass.
 
\subsection{Dressed Ladder BSE}

In this subsection we finally discuss numerical solutions of the BSE 
including the dressing of the propagators. We introduce the dressing by 
two steps, switching it on first only for the exchanged particle and in 
the second step also for constituents. We should point out that since 
the solution of the DSEs in the bare vertex approximation (by which we 
dress the propagators) breaks down for coupling constants larger that 
$\tilde{\lambda}_{crit}= 1.5$  we can consider to only rather weakly 
bound states: for the bare BSE $\tilde{\lambda}= 1.5$ corresponds to 
$\eta=\sqrt{P^2}/(2m)\simeq 0.78$.  

The propagator dressing of the exchanged particle in the one loop 
approximation  was already considered in the Ref. \cite{KUSAK3}. We go 
beyond the one loop approximation and determine the continuum part of 
Lehmann weight  $\sigma_3$ from  the DSE (\ref{SYMB3}) with the same 
value of the coupling constant. That is, we insert into  Eq. 
(\ref{majn} ) the dressed kernel 
\be   \label{opice}
G_3(p-q)=\int_0^\infty d\omega \frac{\tilde{\sigma}_3(\omega)}{(p-q)^2-\omega+i\epsilon}
\ee
with the pole situated at $m_3=m/2$. As noted above, the constituent 
propagators are at this stage left undressed in the BSE, although they 
all self-energies have been taken fully into account in DSEs. The  
integration over $\omega$ (\ref{opice}) in the BSE kernel was performed 
using Gaussian quadrature with 16 points. Including the kernel self energy 
slightly decreases (by at most few per cent) the mass of the bound state,
even for $\tilde{\lambda}\simeq\tilde{\lambda}_{crit.}$.       

Let us point out, that the  kernel of Eq. (\ref{majn}) in the dressed 
ladder kernel approximation is free of any singularities. The accuracy 
of the numerical solution is comparable to the the bare ladder case. 
For example, we $\tilde{\lambda}=0.734$ for $\eta=0.95$  for the grid 
of $32*32$ points and  $\lambda=0.749(0.752)$ for the grid of 
$64*64(96*96)$, the convergence is similar to to the case of bare 
ladder (see the Tab.1). The extrapolated (to very large grid) values of 
$\lambda's$  for fractional binding $\eta=0.999, 0.99, 0.97, 0.95$ are 
showed in Fig.\ref{figBSElam}. 

In the next step  we have included the self energies of the 
constituents. As we shall see the effect is relatively small for  
$\tilde{\lambda} << \tilde{\lambda}_{crit}$, but increases rapidly as 
$\tilde{\lambda} \rightarrow 1$. As in the previous case the Lehmann 
weights have been calculated from the DSEs solved for the same value of 
the coupling. As the first guess we have used the solution of the BSE
linearized in $\tilde{\sigma}(\alpha)$, i.e. with only one propagator dressed.
This guess is rather close to the exact solution for $\eta \leq 0.9$.

The constituent particle in a weakly bound system ($\eta\simeq 1$)   
live  near their mass shell. Therefore, one can naively assume 
that the values of coupling for such weakly bound state should not be 
strongly affected by dressing of constituent propagators. For deeper 
bound states we have found that the effect of the dressing of 
constituents is much larger than that due to the dressing of the kernel 
(exchanged particle) Fig.\ref{figBSElam}. The couplings for fully 
dressed BSE are not determined with the same high accuracy as those for 
the bare ladder BSE, since the grids are not optimalized for very different
ratios of $\alpha$'s which appear in the kernel of the BSE.

\subsection{Charge form factor}

Various form factors are extensively studied in scalar theories like 
Wick-Cutkosky model (see for example \cite{AMGHAR}). In these studies
the dependence of the form factor on the binding and on the range of
the ``strong'' interaction has been considered, therefore we perform
a similar calculations in our formalism.  

In the approach adopted in this paper (employing the spectral 
representations in the Minkowski space), the bound states masses and 
corresponding vertex functions can be obtained with good accuracy and 
in reasonable CPU time. Unfortunately, the calculation of the scalar 
form factor as outlined in appendix B leads to more complicated results 
(\ref{iq},\ref{gqres}). Even if one would be able to perform 
analytically all additional Feynman integrations, the formula 
(\ref{gqres}) still involves the four-dimensional integration over the 
spectral variables. We are taking also the integrals over four Feynman 
variables numerically with the help of Gauss-Legendre quadrature, 
taking the number of points for each of them equal to the one for 
spectral variable $z$. Since relatively small number of integration 
points (from 16 to 40) was taken for each integration, the presented 
results have to be viewed rather as an estimate of form factor 
behavior. One can always refine the grids at the  expense of longer CPU 
time. 

We have also compared our results to choose obtained in the Gross 
(spectator) formalism, choosing the "scalar deuteron" parameters (see 
\cite{ORDEN}). In analogy with the real deuteron, the parameters are 
chosen as: $m_3/m=138/938.9 , \eta=(2*938.9-2.3)/2*938.9 \simeq 
0.9988$. The bound state vertex functions were found by solution of the 
Gross and BS equations. All phenomenological form factors introduced in 
\cite{ORDEN} have been "switched off" (the limit $\Lambda's\rightarrow 
\infty $ are taken in the ``strong'' form factors) when calculating the 
Gross wave function and the bound state current.  The e.m\ form factors 
were calculated in the spectator RIA and as described in appendix B 
(using the grid $32^8$), respectively. 
 
The form factors for several bound states listed in the Tables 
1 and 3 are presented in Figs.\ref{figGfirst} and \ref{figGsecond}, 
respectively. In Fig.\ref{figGfirst} the ratio of the exchanged and 
constituent mass is kept constant $m_3/m= 0.5$ and the mass of the 
bound state $M=\sqrt{P^2}$ is varied. In agreement with physical 
expectations one sees that as the bound state become more tight the 
elastic form factor increases. (For the infinitely bound point system 
is should be equal to unity, even our deepest bound state are of course 
still only approaching this value.) We put into  the same plot  for 
comparison  also the scalar deuteron result. How the form factror 
changes with the mass of the exchanged particle is shown in 
Fig.\ref{figGsecond}. We included several states for which $M=m$, 
bound by the one-boson-exchange potential of the range $r=1/m_3$, which 
is varied. Two other systems, first with $\eta=0.4; m_3=0.25$ and the 
scalar deuteron, are added for comparison.  From both figure we can 
conclude that the behavior of form factor is  determined by the 
strength of the interaction and its range rather differently for 
various $M$. The range of the interaction is more significant for a 
weaker ones. This agrees with conclusions of ref. \cite{THEUSL}, which 
have also compared our results for small exchanged mass $\eta=0.8, 
 m_3=0.15m$ with Wick-Cutkosky model prediction  for 
$\eta=0.784, m_3=0$ (Tables 2 and 1 of \cite{THEUSL}, respectively), 
and found only slight difference in the range $Q^2=(0,100m^2)$.

\section{Concluding Remarks}

The  spectral representation was employed for solving the Bethe-Salpeter 
equation in (3+1) Minkowski space.  
The new analytical formula for the integral equation kernel has been derived. 
 
The method is efficient  solving both the bare and dressed ladder BSE. 
Solution of Dyson-Schwinger equations for propagators leads to 
appearance of critical value of the coupling constant, beyond which the 
solution collapses. This restricts substantially the region in which 
the effects of dressing can be studied. Since the coupling is rather 
weak, the dressing leads only to moderate decrease of the bound state 
masses: even close to the critical value of the coupling the fractional 
binding of the bound state of the dressed BSE is smaller than the 
corresponding one for the bare BSE by at most 15 per cents. As an 
example of application of the obtained vertex functions, we calculated 
the elastic electromagnetic form factor. 

To further develop the method, it would be interesting to extend it to 
more complicated BS kernel: trying to include the cross boxed 
contributions, $s$ and $u$ channel interactions etc. It is already 
known that the ``spectral'' approach used here is  suitable even for 
more complicated systems, for scalar QED see Ref. \cite{FULTON}. One of 
our future goals is to manage the complication due to fermionic degrees 
of freedom. 

\section*{Acknowledgments}

The calculation of scalar deuteron electromagnetic form factor in 
quasipotential approximation was part of V.\v{S}auli's Diploma thesis
and a preliminary version of form factor calculations was reported at
\cite{SAULI}. 
This research was supported by GA \v{C}R under Contract n. 202/00/1669. 

\newpage

\appendix

\section{Kernel Functions}

In this appendix the real integral equation for the BS vertex weight is 
derived in detail. The PTIR form for scalar bound-state vertex reads 
\be  \label{jedna}
\Gamma(p,P)= \int\limits_{-1}^1 dz\,
\int\limits_{\alpha_{min}(z)}^{\infty}d\alpha\, \, 
\frac{\rho^{[n]}(\alpha,z)}{[F(\alpha,z;p,P)]^n} \quad ,
\ee
where  $\rho^{[n]}(\alpha,z)$ is the real PTIR weight function
for the bound state vertex function, and $n$ is a dummy parameter.
The function $F$ is given by \cite{NAKANATO}
\be
F(\alpha,z;p,P)=\alpha-(p^2+zp\cdot P+P^2/4)- i \epsilon=
               \alpha- f(p,P,z) - i \epsilon \quad .
\ee
The support of $\rho^{[n]}(\alpha,z)$ can be determined in general
(see \cite{NAKANATO}) for arbitrary interaction. In our case of
one-boson-exchange interaction kernel one gets a bit higher 
$\alpha_{min}$ \cite{KUSAK3}. We discuss our treatment of the
lower bounds below.
 
The following procedure is straightforward but a bit exhaustive. The 
rhs of the BSE (\ref{bse}) with the kernel given by the exchange of 
the single (dressed) particle $\Phi_3$ has to be re-written in the form 
allowing to extract the integral equation for the spectral function 
$\rho^{[n]}(\alpha,z)$. The integrand contains the two ``constituent'' 
propagators, the denominator $F(\alpha,z,k,P)$ from the spectral 
representation of $\Gamma(k,P)$ and the propagator of the exchanged 
particle (all other factors will be skipped for a while for the sake of 
briefness). 

Using the  Feynman parameterization technique we first write 
\bea
 && D(k+ P/2;\alpha_1) D(-k+P/2;\alpha_2)=
 \frac{1}{2} \int\limits_{-1}^1 
 \frac{d \eta}{[M^2- f(k,P,\eta)- i \epsilon]^2} \, , \nonumber \\
 && M^2(\eta)= \frac{\alpha_1+\alpha_2}{2}+ \frac{\alpha_1-\alpha_2}{2}\, \eta \, .
\eea
Next the denominator of (\ref{jedna}) is added:
\bea
&&\frac{D(k+ P/2;\alpha_1) D(-k+P/2;\alpha_2)}{[F(\alpha,z;k,P)]^n}=
\frac{\Gamma(n+2)}{2\Gamma(n)}\,  \int\limits_{-1}^1 d \eta
\int\limits_0^1 d t 
\frac{(1-t)t^{n-1}}{[R - f(k,P,z')- i\epsilon]^{n+2}} \, , \nonumber \\
&&R= \alpha t + (1-t)M^2 \, , 
\eea
where $z'= t z+ (1-t)\eta$. Now, we include the propagator of the exchanged 
particle, the integral over $d^4 k$ and factors $i g^2$, defining: 
\bea
&&I_{DDDF}= i g^2\,  \int \frac{d^4 k}{(2\pi)^4} 
 \frac{D(k+ P/2;\alpha_1) D(-k+P/2;\alpha_2) D(p-k;\alpha_3)}{[F(\alpha,z,p,P)]^n} \\
&&\hspace*{2.0truecm}
 =- i g^2\, \frac{\Gamma(n+3)}{2\Gamma(n)}\,  \int\limits_{-1}^1 d \eta
\int\limits_0^1 d t\, (1-t)t^{n-1}\int\limits_0^1 d x\, x^{n+1}\, I_k \, , \nonumber\\
&&I_k= \int \frac{d^4 k}{(2\pi)^4} 
\left[ -k^2+ 2k\cdot Q - (1-x)p^2- \frac{x}{4}P^2
+(1-x)\alpha_3+ x R- i\epsilon \right]^{-(n+3)} \nonumber \\
&& \hspace*{2.0truecm}
= \frac{i}{(4\pi)^2}\frac{\Gamma(n+1)}{\Gamma(n+3)}\frac{1}{x^{n+1}(1-x)^{n+1}}\,
\frac{1}{[ A - f(k,P,z')- i\epsilon ]^{n+1}} \nonumber \\
&& A= \frac{R}{1-x}+ \frac{\alpha_3}{x}- \frac{x}{(1-x)}\, S \, ,
\eea
where $Q= (1-x)p- x z' P/2$ and $S= \frac{1- z'^2}{4}P^2$. Since $z'$ lies in the
interval $<-1,+1>$, $0\leq S < (m_1+ m_2)^2/4$. Interchanging the 
integrals over $\eta$ and $t$ with the help of: 
\bea && \int\limits_{-1}^{1}d\eta\int\limits_0^1 dt= 
\int\limits_{-1}^{1}d z'\left[\int_0^{T_+}\frac {dt}{1-t}\, 
\Theta(z-z') +\int_0^{T_{-}}\frac {dt}{1-t}\, \Theta(z'-z)\right] \, , 
 \nonumber\\ 
&&T_\pm= \frac{1 \pm z'}{1 \pm z} \quad \quad \mbox{and} \quad \quad 
z'= t z+ (1-t)\eta \, , 
\eea 
and introducing $\lambda= g^2/(4\pi)^2$ we get 
\be
I_{DDDF}= \lambda \, \frac{n}{2}  \int\limits_{-1}^1 d z'
 \int\limits_0^1 \frac{d x}{(1-x)^{n+1}} \sum_{s=\pm} \Theta(s(z-z'))
  \int\limits_0^{T_s} \frac{d t\, t^{n-1}}{[F(A,z',p,P)]^{n+1}} \, , 
\ee
Let us separate the $t$ dependence of $F(A,z';p,P)$. First we substitute
for $\eta$ into the definition of $M^2$, which yields 
\be
(1-t) M^2= -t \left(\frac{\alpha_1+\alpha_2}{2}+ \frac{\alpha_1-\alpha_2}{2} z \right) 
+ \frac{\alpha_1+\alpha_2}{2}+ \frac{\alpha_1-\alpha_2}{2}\, z' \, . \nonumber
\ee
Next, we introduce the notation (indicating explicitly the $t$-dependence of 
$A$ and $R$):
\bea  \label{pruser}
&&A(t)\equiv \frac{R(t)}{1-x} +\frac{\alpha_3}{x} -\frac{x}{1-x}\, S=
 \frac{R(t)-S}{1-x} +\frac{\alpha_3}{x}+ S \, ,\\
&&R(t)\equiv \alpha t+ (1-t) M^2= J(\alpha,z)t+ 
  \frac{\alpha_1+\alpha_2}{2}+\frac{\alpha_1-\alpha_2}{2}\, z' \\
&&J(\alpha,z)\equiv\  
\alpha-\frac{\alpha_1+\alpha_2}{2}-\frac{\alpha_1-\alpha_2}{2}\, z \, ,
\eea
in which the $t-$dependence of $F(A(t),z',p,P)$ reads
\be
 F(A(t),z';p,P)= \frac{J(\alpha,z)}{1-x}\, t+  F(A(0),z';p,P) \, .
\nonumber 
\ee
Since t-dependence of $F(A(t),z';p,P)$ is linear, the integral over $t$ can be taken:
\be
\int \frac{d t\, t^{n-1}}{ F(A(t),z';p,P)^{n+1}}= 
\frac{t^n}{n\, F(A(0),z';p,P)\, [F(A(t),z';p,P)]^n} \, .
\nonumber
\ee
and hence
\be
I_{DDDF}= \frac{\lambda}{2}\, \int\limits_{-1}^1 d z'
 \int\limits_0^1 \frac{d x}{(1-x)^{n+1}} \sum_{s=\pm} 
 \frac{\Theta(s(z-z'))\, T_s^n}{F(A(0),z';p,P)\, [F(A(T_s),z';p,P)]^n} \, .
\label{Idddf}
\ee
Using this result, the BSE can be written as follows:
\bea
&& \int\limits_{-1}^1 dz' \,
\int\limits_{\alpha_{min}(z')}^{\infty}d\alpha' \, \,
\frac{\rho^{[n]}(\alpha',z')}{[F(\alpha',z',p,P)]^n} =
 \left[\int\limits\sigma\right]^3
\int d \alpha \int d z \rho^{[n]}(\alpha,z)\, I_{DDDF} \, .\, 
\label{bsei} \\
&&\left[\int\limits\sigma\right]^3\equiv
\int\limits_{\alpha_{1,min}}^{\infty} d\alpha_1 \, \tilde{\sigma}_1(\alpha_1)
\int\limits_{\alpha_{2,min}}^{\infty} d\alpha_2 \, \tilde{\sigma}_2(\alpha_2)
\int\limits_{\alpha_{3,min}}^{\infty} d\alpha_3 \, \tilde{\sigma}_3(\alpha_3) \, ,
\eea
where $\tilde{\sigma}({\alpha_i})$ are the Lehmann functions of the 
dressed propagators, reducing to the $\delta$-function if the dressing 
is neglected. Obviously, the rhs is still not quite in the desired form, 
both $F(A(0),z',p,P)$ and $F(A(T_s),z',p,P)^n$ are functions of momenta 
$p$ and $P$. 

Below we re-write the kernel as a sum of several fractions, which after 
substitution $\alpha'= A(T), T= 0, T_{\pm}$ would allow to use the 
uniqueness theorem \cite{NAKANATO} and extract the BSE in the spectral 
form. This is possible only if the integrals over $\alpha'$ on both 
right and left hand sides of \ref{bsei} are taken over the same 
intervals. To show this it is first necessary to prove that $R(T)- S > 
0$, for $T= 0, T_{\pm}$, since the functions $A(T) \rightarrow + 
\infty$ for $x \rightarrow 0$ and $x \rightarrow 1$ they have on for $0 
< x < 1$ the minimum equal to:   
\be
 A_{min}(T) = \left( \sqrt{R(T)- S} + m_3 \right)^2 + S =
  R(T)+ m^2_3+ 2 m_3\, \sqrt{(R(T)-S)}\, .  
\label{atbound}
\ee
next, one has to show that these lower bounds are not in conflict with
the lower bound for $\alpha'$ (and the same lower bound for $\alpha$).
It is simple for the undressed equal mass case, when the condition above
taken for $T=0$ actually defines the lower bound for $\alpha$ in
the following form
\be
\alpha \geq \left( \sqrt{m^2- S(z')} + m_3 \right)^2 +S(z') \, ,
\ee
and the same for $\alpha, z \rightarrow \alpha', z'$. Since $R(T_\pm)$
depend also on $\alpha$ and $z$ and since for the equal mass case
$R(T_\pm) \geq R(0)$, the next two constrains clearly conform with
the lower bound for $\alpha'$ and one can extract from them the upper bound
for the integration over $\alpha$:
\be
 \alpha \leq m^2+ \frac{1}{T_\pm}\,  
 \left[ \left( \sqrt{\alpha'- S(z')}- m_3 \right)^2 +S(z')- m^2 \right] \, , 
\ee
(compare to eq.(A6) of \cite{KUSAK3}, where one bracket seems to be 
misplaced). For the unequal mass case or for the dressed propagators 
this analysis is much more complicated, mostly due to the fact that now  
for some combination of parameters it can occur $R(T_\pm) < R(0)$. The 
necessary condition $R(T)-S > 0$ can again be proven in this case 
(though after much longer algebra), ensuring that the common lower 
bound for $\alpha$'s exists. But we could not resolve the conditions 
(\ref{atbound}) analytically, they are treated numerically.   

Now, let so go back to eq.\ ({\ref{bsei}}) and proceed by considering 
first the simple case of the symmetric Wick-Cutkosky model. 

\subsection{The Wick-Cutkosky Model}

In this model the constituents have the same masses $m_1=m_2=m$ and the 
mass of the exchanged particle is zero (all propagator dressings are 
neglected). Then for the S-wave ground state vertex function, the 
spectral function depends only on variable $z$ and the denominator 
enters with power $n=1$ (\ref{wcm}): 
\be
\rho^{[n]}(\alpha,z) \rightarrow \delta (\alpha- m^2) \rho (z) \, , 
\ee
and in $I_{DDDF}$ we replace
\bea
&& n \rightarrow 1 \, , \quad \quad M^2 \rightarrow m^2 \, , 
\quad \quad R(t) \rightarrow m^2 \, ,
 \quad \quad J(\alpha,z) \rightarrow 0 \, , \nonumber\\
&& A(t) \rightarrow A= \frac{m^2}{1-x} - \frac{x}{1-x}\, S \, , \nonumber\\
&& I_{DDDF}\rightarrow \frac{\lambda}{2}\, \int\limits_{-1}^1 d z'
  \sum_{s=\pm} \Theta(s(z-z'))\, T_s \,
 \int_0^1 \frac{d x}{[m^2- x S- (1-x)f(p,P,z')- i\epsilon]^2}   \, . 
\nonumber
\eea
Taking the integral over $x$ we find (recall that for equal masses 
$0 \leq S < m^2$):
\be
 I_{DDDF}\rightarrow \frac{\lambda}{2}\, \int\limits_{-1}^1 d z'
 \sum_{s=\pm} \frac{\Theta(s(z-z'))\, T_s}{m^2-S}\,
 \frac{1}{m^2+ f(p,P,z')- i\epsilon} \, . \nonumber
\ee
Comparing both sides of the BSE and using the uniqueness theorem 
\cite{NAKANATO}, we get the well-known (see e.g.\ \cite{Nakan}) 
integral equation for $\rho^{[1]}(z)$: 
\bea
\rho (z') &=& \lambda \int\limits_{-1}^1 d z V^{[1]}(z',z)\, \rho (z) \, , \\
V^{[1]}(z',z) &=& \sum_{s=\pm}\, \frac{\Theta(s(z-z'))\, T_s}{2(m^2-S)} \, . \nonumber
\eea
Although its solution is known analytically even for  exited states, 
the energy spectrum still has to be found numerically (up to the 
$P^2=0$ corresponding to $\lambda=2$). For the purpose of numerical 
treatment, this equation is usually rewritten in the form: 
\bea   \label{wcmsolve}
\frac{\rho(z')}{\lambda}&=& V_0(z')-\int\limits_{-1}^{1}dz \rho(z)
V_{\pm}(z',z)
\nn \\
V_0(z')=\frac{1}{2(m^2-S)} \quad &;& \quad
V_{\pm}(z',z)=
\frac{\frac{z'-z}{1-z}\Theta(z'-z)+\frac{z-z'}{1+z}\Theta(z-z')}
{2(m^2-S)} \, \, \nn
\eea
where the temporary auxiliary normalization condition $\int dz \rho 
(z)=1$ was imposed (i.e., if used in further application, the vertex 
function would have to be re-normalized in accordance to 
(\ref{normalization})). 
 
\subsection{The BSE for $n=1$}

We will now bring eqs.\ ({\ref{Idddf},\ref{bsei}) to the desired form
for the particular choice $n=1$. For the numerical solution this value is
not the most suitable one. But since the formal manipulations are  in
this case the simplest, we treat it first for methodical reasons.

For $n=1$ the integrand of (\ref{Idddf}) can be decomposed into the
sum of simpler fractions with the help of
\be
 \sum_{s=\pm} 
 \frac{\Theta(s(z-z'))\, T_s}{F(A(0),z';p,P)\, F(A(T_s),z';p,P)} =
 \frac{1-x}{J(\alpha,z)}\, \sum_T \frac{1}{F(A(T),z';p,P)} \, , 
\label{decomp}
\ee
where we have introduced a shorthand notation
\be
\sum_T f(T)= f(0)- \theta(z-z')f(T_+)- \theta(z'-z)f(T_-) \, .
\ee
Notice that the lhs of (\ref{decomp}) is nonsingular for 
$J(\alpha,z)=0$, so when this happens the rhs behaves like $0/0$, which 
calls for some caution in the numerics. So, the integral $I_{DDDF}$ 
now reads 
\be
 I_{DDDF}(n=1)= \frac{\lambda}{2J(\alpha,z)}\,
 \sum_T  \int\limits_0^1 \frac{d x}{(1-x)F(A(T),z';p,P)} \, .
\ee
In the last step we introduce the spectral variable $\alpha'= A(T)$
and use the dependence of $A(T)$ on $x$ to convert the integration over
$x$ into integral over $\alpha'$. Picking up explicitly the $x$-dependence
of $A(T)$ we can write:
\bea
g(x)&\equiv& A(T)=  \frac{R(T)-S}{1-x} +\frac{\alpha_3}{x}+ S \, , \nonumber\\
g'(x)&=& \frac{R(T)-S}{(1-x)^2}- \frac{\alpha_3}{x^2}=
\frac{1}{1-x}\left(A- \frac{\alpha_3}{x^2}- S\right) \, , \nonumber\\
\delta (\alpha'- g(x))&=& \sum_{i=\pm} \frac{1}{|g'(x_i)|}\, \delta(x-x_i) \, ,
\quad  \quad  g(x_\pm)= \alpha' \, , \nonumber\\
x_{\pm}(T)&=& \frac{\alpha'-\alpha_3-R(T) \pm \sqrt{D}}{2(\alpha'_S)} \, ,
 \quad
D= (R(T)-\alpha'+\alpha_3)^2- 4\alpha_3(R(T)-S) \, . \nonumber\\
g'(x_\pm)&=& \left( g(x_\pm)-\frac{\alpha_3}{x_\pm^2}-S \right) \frac{1}{1-x_\pm}= 
\frac{E(x_\pm,S,\alpha')}{1-x_\pm} \, ,\nonumber\\
E(x_\pm,S,\alpha')&=& \alpha'- \frac{\alpha_3}{x_\pm^2} -S \, .
\eea
With the help of these relations we get
$$
\sum_T  \int\limits_0^1 \frac{d x}{(1-x)F(A(T),z';p,P)}=
\int\limits_{\alpha_{min}^\infty} \frac{d \alpha'}{F(\alpha',z';p,P)}\,
\sum_T  \, \sum_{i=\pm} 
\frac{\theta(x_i(T))\theta(1-x_i(T))\theta(D)}{|E(x_i(T),S,\alpha')|}
\, .
$$
Using this result in (\ref{bsei}), one gets from the uniqueness theorem 
the integral equation for the BSE structure function is obtained: 
\bea
 \rho^{[1]}(\alpha',z') & =& \lambda\, \left[\int\limits\sigma\right]^3
\int d z \int d \alpha \, V^{(1)}(\alpha,z,\alpha',z')\, 
\rho^{[1]}(\alpha,z)\,  , 
\label{bse1} \\
 V^{(1)}(\alpha,z,\alpha',z')&=& \frac{1}{2J(\alpha,z)}\,
 \sum_T  \, \sum_{i=\pm} 
 \frac{\theta(x_i(T))\theta(1-x_i(T))\theta(D)}{|E(x_\pm(T),S,\alpha')|}
 \, .
\eea
Notice, that $x_\pm(T)$ depend on $\alpha', z'$ (which are fixed from the lhs),
$\alpha_1, \alpha_2, \alpha_3$ (which are fixed when the dressing is neglected),
$P^2$ in $S$ (which is given by the binding energy of the system) and for
$T\neq 0$ also on $\alpha, z$ (through $J(\alpha,z)$ in $R(T_\pm)$). This allows
to recast the integral equation into the form more convenient for numerical
treatment. 

\subsubsection{BSE without propagator dressing (for $n=1$)}

If the propagator dressing is omitted, $\alpha_i \rightarrow m_i^2$ and
the corresponding BSE for the spectral function reads
\bea
\rho^{[1]}(\alpha',z')&=& \lambda\, \int d z \int d \alpha \,  
 V^{[1]}(\alpha',z',\alpha,z) \rho^{[1]}(\alpha,z)  \,  , \nonumber\\
 V^{[1]}(\alpha',z',\alpha,z)&=&  \frac{1}{2J(\alpha,z)}\, \sum_T  \, \sum_{i=\pm}
 \frac{\theta(x_i(T))\theta(1-x_i(T))\theta(D)}{|E(x_i(T),S,\alpha')|} 
\nonumber
\eea
As mentioned above, for $T=0$ the kernel depends on $\alpha$ and $z$ only
through $J(\alpha,z)$, hence it is convenient to pick up this case
from the sum over $T$ and re-scaling 
\bea
\rho^{[1]}(\alpha,z)&=& J(\alpha,z)\, \tilde{\rho}^{[1]}(\alpha,z) \, , \nonumber\\
 V^{[1]}(\alpha',z',\alpha,z)&=& \frac{J(\alpha',z')}{J(\alpha,z)}\, 
\tilde{V}^{[1]}(\alpha',z',\alpha,z) \, , \nonumber 
\eea
and imposing the auxiliary normalization
\be
  \int d z \int \alpha \, \tilde{\rho}^{[1]}(\alpha,z) = 1 \, , 
\nonumber
\ee
rewrite the BSE in the non-homogenous form:
\bea 
\label{baren1solve}
\tilde{\rho}^{[1]}(\alpha',z')&=& \lambda\, \tilde{V}_0^{[1]}(\alpha',z')-
 \lambda\,  \int d z \int \alpha \, \sum_{s=\pm} 
 \tilde{V}_s^{[1]}(\alpha',z',\alpha,z)\, \tilde{\rho}^{[1]}(\alpha,z) \, , \\
\tilde{V}_0^{[1]}(\alpha',z')&=& \frac{1}{2J(\alpha',z')}\, \sum_{i=\pm} 
 \left[ \frac{\theta(x_i(T))\theta(1-x_i(T))\theta(D)}{|E(x_\pm(T),S,\alpha')|} \right]_{T=0}
 \, ,
\nonumber\\
\tilde{V}_s^{[1]}(\alpha',z',\alpha,z)&=& \frac{\theta(s(z-z'))}{2J(\alpha',z')}\, 
 \sum_{i=\pm} \left[ 
  \frac{\theta(x_i(T))\theta(1-x_i(T))\theta(D)}{|E(x_\pm(T),S,\alpha')|} 
\right]_{T=T_s}   \, .  \nonumber     
\eea
The $\theta$ functions in the kernel impose the proper bounds on $\alpha$
and ensure that the rhs contributes only for $\alpha'$ from the support of
$\tilde{\rho}^{[1]}(\alpha',z')$.

\subsection{BSE for $n=2$}

Now, we would first describe the necessary modifications for the choice 
$n=2$ which was used in actual numerical calculations in this paper. Going 
back to (\ref{Idddf}) we can for this case write: 
\bea
 && \sum_{s=\pm} 
 \frac{\Theta(s(z-z'))\, T_s^2}{F(A(0),z',p,P)\, F(A(T_s),z',p,P)^2} = 
 \nonumber\\
 &&\hspace*{3.0truecm} 
 \frac{1-x}{J(\alpha,z)}\, \sum_T 
 \left[ \frac{T}{[F(A(T),z';p,P)]^2}+ 
 \frac{1-x}{J(\alpha,z)}\, \frac{1}{F(A(T),z';p,P)} \right] \, . 
\nonumber
\eea
Then, the integral $I_{DDDF}$ can be cast into form:
\bea
 I_{DFFF}&=& \frac{\lambda}{2 J(\alpha,z)}\,
 \int_{-1}^1 d z' \int_0^1 \, \frac{d x}{(1-x)^2}\,
\sum_T 
 \left[ \frac{T}{[F(A(T),z';p,P)]^2}+ 
 \frac{1-x}{J(\alpha,z)}\, \frac{1}{F(A(T),z';p,P)} \right] \nonumber\\
 &=& \frac{\lambda}{2 J(\alpha,z)}\, \int_{-1}^1 d z' 
 \sum_T \, \frac{d x}{[F(A(T),z';p,P)]^2}\,
 \left[ \frac{T}{(1-x)^2}- \frac{\ln{(1-x)}}{J(\alpha,z)}\frac{d A(T)}{d x} 
 \right] \nonumber\\
  &=& \frac{\lambda}{2 J(\alpha,z)}\, \int_{-1}^1 d z' 
  \int_{\alpha_{min}}^{\infty} \frac{d \alpha'}{[F(\alpha',z';p,P)]^2}\,
   \sum_T \,  \sum_{i=\pm} \, \theta(x_i) \theta(1-x_i) \theta(D)\nonumber\\
&&  \hspace*{2.0truecm} 
   \left[ \frac{T}{(1-x_i)|E(x_i,S,\alpha')|}- 
  \frac{\mbox{sgn}(E(x_i,S,\alpha')\ln{(1-x_i)}}{J(\alpha,z)} 
 \right] \, , \nonumber
\eea  
where we have first integrated the second term of the sum over $T$ by 
parts (to increase the power of $F(A(T),z';p,P)$, the boundary term 
vanishes when $x\rightarrow 0,1$) and then introduced the integration 
over $\alpha'$ as in the previous section. 

Now, we can use the uniqueness theorem of PTIR \cite{NAKANATO} and identify
the BS weight function on the right-hand side of BSE: 
\bea   \label{sest}
\rho^{[2]}(\alpha',z')&=& \lambda\, \int\limits_{-1}^{1} dz 
\int\limits_{-\infty}^{\infty} d \alpha \, 
V^{[2]}(\alpha',z';\alpha,z) \rho^{[2]}(\alpha,z) \,  , \\
V^{[2]}(\alpha',z';\alpha,z)&=&
\left[\int \sigma\right]^3
\sum_T \sum_{i=\pm}
\frac{\theta(x_i)\theta(1-x_i)\theta(D)}{2J(\alpha,z)^2} \nonumber\\
&& \hspace*{0.5truecm}
\left\{\frac{T J(\alpha,z)}{(1-x_i)|E(x_i,S,\alpha')| }
- \mbox{sgn}\left(E(x_i,S,\alpha')\right)\ln(1-x_i) \right\}.
\eea
Before treating the general case with fully dressed propagator we consider 
in the next subsection the pure ladder BSE.

\subsubsection{BSE without propagator dressing (for $n=2)$}

In this case we can proceed in a way very similar to the undressed BSE
for $n=1$, only with different re-scaling factor
\bea
\label{redef}
\rho^{[2]}(\alpha,z)&=& J(\alpha,z)^2\, \tilde{\rho}^{[2]}(\alpha,z) \, , \nonumber\\
 V^{[2]}(\alpha',z',\alpha,z)&=& \frac{J(\alpha',z')^2}{J(\alpha,z)^2}\, 
\tilde{V}^{[2]}(\alpha',z',\alpha,z) \, . \nonumber 
\eea
Imposing the auxiliary normalization
\be
  \int d z \int \alpha \, \tilde{\rho}^{[2]}(\alpha,z) = 1 \, , 
\nonumber
\ee
we can write
\bea 
\label{baren2solve}
\tilde{\rho}^{[2]}(\alpha',z')&=& \lambda\, \tilde{V}_0^{[2]}(\alpha',z')-
 \lambda\,  \int d z \int \alpha \, \sum_{s=\pm} 
 \tilde{V}_s^{[2]}(\alpha',z',\alpha,z)\, \tilde{\rho}^{[2]}(\alpha,z) \, , \\
\tilde{V}_0^{[2]}(\alpha',z')&=& \left. - \sum_{i=\pm}
\frac{\theta(x_i)\theta(1-x_i)\theta(D)}{2J(\alpha',z')^2}
{\mbox{sgn}}\left(E(x_i,S,\alpha')\right)\ln(1-x_i)\right|_{x_i=x_i(T=0)} \, \nonumber\\ 
\tilde{V}_s^{[2]}(\alpha',z',\alpha,z)&=& \frac{\theta(s(z-z'))}{2J(\alpha',z')^2}\, 
  \sum_{i=\pm} \theta(x_i)\theta(1-x_i)\theta(D)\, \nonumber\\
&& \hspace*{0.5truecm}
\left\{\frac{T_s J(\alpha,z)}{(1-x_i)|E(x_i,S,\alpha')| }
-{\mbox{sgn}}\left(E(x_i,S,\alpha')\right)\ln(1-x_i)\right\}_{x_i=x_i(T_s)}
  \, .
\nonumber
\eea 
 
\subsubsection{Dressed ladder BSE for $n=2$}
 
When the all self-energies are taken into account  the function 
$J(\alpha,z)$ do not factorize and numerically convenient redefinition 
(\ref{redef}) can not be  used. Nevertheless, we can still separate from
the integral over spectral variables $\alpha_i$ and the sum over $T$ the
term which depends on $\alpha$ and $z$ only through $J(\alpha,z)$, namely
the part for which $\tilde{\sigma}_i \rightarrow \delta(\alpha_i -m_i^2)$ and
$T=0$. Explicitly, using the notation
\bea
 {\cal V}(\alpha',z',\alpha,z,T,\alpha_1,\alpha_2,\alpha_3)&=&
 \sum_{i=\pm} \frac{\theta(x_i)\theta(1-x_i)\theta(D)}{2J(\alpha,z)^2}\, \nonumber\\
&& \hspace*{-1.0truecm}
\left\{\frac{T J(\alpha,z)}{(1-x_i)|E(x_i,S,\alpha')| }
-{\mbox{sgn}}\left(E(x_i,S,\alpha')\right)\ln(1-x_i)\right\}_{x_i=x_i(T)}
  \, ,
\nonumber  
\eea 
and imposing the normalization condition
\be  \label{ncecko}
1= \int\limits_{-1}^{1}dz\int\limits_{-\infty}^{\infty}
d\alpha \frac{\rho^{[2]}(\alpha,z)}{J(\alpha,z)^2} \, ,
\nonumber
\ee
the BSE can be re-written as
\bea
\rho^{[2]}(\alpha',z')&=& \lambda\, V_0^{[2]}(\alpha',z')+
 \lambda\,  \int d z \int d \alpha \, \sum_{T} 
 V_T^{[2]}(\alpha',z',\alpha,z)\, \rho^{[2]}(\alpha,z) \, , \nonumber\\
V_0^{[2]}(\alpha',z')&=&  - \frac{1}{2} \int d\alpha_3\, \tilde{\sigma}(\alpha_3)
\sum_{s=\pm} \sum_{i=\pm}
\theta(x_i)\theta(1-x_i)\theta(D)\,
{\mbox{sgn}}\left(E(x_i,S,\alpha')\right)\ln(1-x_i)  \, , \nonumber\\ 
V_T^{[2]}(\alpha',z',\alpha,z)&=& \int d \alpha_3\, \tilde{\sigma}(\alpha_3)
\int d\alpha_1 d\alpha_2\, 
\nonumber\\
&&\left[\delta_1 \delta_2 (1-\delta_{T,0})+ 
\delta_1 \sigma_2+ \sigma_1 \delta_2 + \sigma_1 \sigma_2 \right]\,
{\cal V}(\alpha',z',\alpha,z,T,\alpha_1,\alpha_2,\alpha_3) 
  \, ,
\nonumber
\eea
where in $V_0^{[2]}(\alpha',z')$ we take $x_i=x_i(T=0)$, $\alpha_1=m_1^2$, 
$\alpha_2=m_2^2$ and in the last term $\delta_i= \delta(\alpha_i-m_i^2)$
and $\sigma_i= \sigma(\alpha_i-m_i^2), i=1,2$.
 
\newpage

\newpage
\section{Elastic Electromagnetic Form Factor }

In this Appendix we derive the expression for the elastic 
electromagnetic form factor $G(Q^2)$. The  relevant matrix element  is  
diagrammatically depicted in the Fig.\ref{figGmu} and the starting 
equation for the e.m.\ matrix element is given by (\ref{chargeff}). The 
labeling of momenta corresponds to the Fig.\ref{figGmu}: $q$  is a 
virtual photon incoming momentum and we put $Q^2=-q^2>0$, $P_i=P$ and 
$P_f=P+q$ are the total momenta of bound state in in- and out-state, 
respectively. For simplicity, we will consider the equal mass case:
$m_1^2=m_2^2=m^2$.
 
The form factor $G(Q^2)$ can easily be extracted from its definition (\ref{formf}). 
After  multiplying by $P_i+P_f$ we get
\be  \label{defin}
G(Q^2)=\frac{G_{\mu}(P+q,P)(2P+q)^{\mu}}{e(2P+q)^2}  \quad ,
\ee
The matrix element is evaluated between on-shell states of appropriate 
composite scalar,  i.e.\ $ P^2=(P+q)^2=M^2$ which implies 
\be \label{impo}
2P.q+q^2=0
\ee
The one body current $j_\mu$ reads
\begin{equation}
j^{\mu}(p_f,p_i)=(P+2k+q)^{\mu} \quad .
\end{equation}
Using (\ref{impo}) one can simplify
\bea
(2P+q)^2&=& 4 M^2+ Q^2 \, , \nonumber\\
(2P+q)\cdot((2k+ q+ P)&=& 2k\cdot (2P+q)+ 2M^2+\frac{Q^2}{2} \, \nonumber\\
\frac{(2P+q)\cdot((2k+ q+ P)}{(2P+q)^2}&=& \frac{1}{2}\left( 
1+ \frac{4k\cdot (2P+q)}{4 M^2+ Q^2} \right) \, . \nonumber
\eea
Taking this into account we can write:
\bea 
 G(Q^{2})&=& \frac{i}{2}\, \int\frac{d^4k}{(2\pi)^4 2}\, 
\left(\prod_{i=1}^3 D_i\right)\, 
\left(\bar{\Gamma} \Gamma \right)\,
\left[1+\frac{4k\cdot(2P+q)}{4 M^2+ Q^2} \right] \quad , \label{ppp}\\
\prod_{i=1}^3 D_i&=& D(k+\frac{P}{2};m^2) D(-k+\frac{P}{2};m^2) 
D(k+q+\frac{P}{2};m^2) \, \label{ddd} \, , \\
\bar{\Gamma} \Gamma&=& \bar{\Gamma}(k+\frac{q}{2},P+q) \Gamma(k,P)
\, . \label{GG}
\eea

Now, we will express $G(Q^2)$ in terms of spectral
functions of the bound state vertex functions $\Gamma$, re-writing
first the product of the propagators with the help of the Feynman
parametrization. 

For the product of the propagators one gets:
\bea \label{propazi}
\prod_{i=1}^3 D_{i}&=& D(k+q+\frac{P}{2})\, \,
\frac{1}{2}\int\limits_{-1}^{1}d\eta
\frac{1}{\left[k^2+ \eta k\cdot P +\frac{P^2}{4}-m^2 +i\epsilon\right]^2}
\nn \\
&=&\int\limits_{0}^{1}ds \int\limits_{2s-1}^{1}dy
\frac{1}{[k^2+ \frac{M^2}{4}- m^2-\frac{s}{2}Q^2+ 2s k\cdot q+ y k\cdot P +i\epsilon]^3}
\eea
where the substitution $y=s+ (1-s)\eta$ has been used and relations 
$P\cdot q=Q^2/2$ and $P^2=M^2$ has been replaced used.

Next step we combine PTIR (for $n=2$) of the product of the bound 
state vertex functions. 
\bea  \label{ggkriz}
 &&\int\limits_0^{\infty}d\alpha_1 d\alpha_2
\int\limits_{-1}^{1}dz_{1}dz_{2}\rho^{[2]}(\alpha_1,z_1)\rho^{[2]}(\alpha_2,z_2)
\left\{\int\limits_{0}^{1}dx\frac{\Gamma(4)}{\Gamma(2)\Gamma(2)}
\frac{ x (1-x)}{G^4}\right\}
\nn \\
G&=&k^{2}+ \frac{M^2}{4}+ (z_1+ x(z_2-z_1)) k\cdot P+ x(1+z_2) k\cdot q-
 x(1+z_2)\frac{Q^2}{4}\nonumber\\
 && -(1-x) \alpha_1- x \alpha_2 + i\epsilon \, .
\eea

Making use of the Feynman variable $t$ for matching (\ref{propazi}) 
with the term in large brackets of (\ref{ggkriz}), the relation for
the form factor (\ref{ppp}) can be rewritten as 
\bea  \label{ooo}
I(Q^2)&=& -\frac{i}{2}\frac{\Gamma(7)}{\Gamma(3)}\,
\int\limits_{0}^{1}ds\int\limits_{2s-1}^{1}dy
\int\limits_{0}^{1}dx\,  x (1-x)\, \int\limits_{0}^{1}dt\, t^3(1-t)^2
\nonumber\\
&& \int \frac{d^4k}{(2\pi)^4}\, 
\frac{\left[1+4\frac{(2P+q).k}{(2P+q)^{2}}\right]}
{\left[c-k^2 -k\cdot (aP+bq)-i\epsilon\right]^7}
\nn \\
a&=& t [z_1+x(z_2-z_1)] + (1-t)y \, ,
\nn \\
b&=& tx(1+z_2)t+ 2s(1-t) \, ,
\nn \\
c&=& b\, \frac{Q^2}{4}- \frac{M^2}{4}+ (1-t)m^2+ t(1-x)\alpha_{1}+ t x \alpha_{2} 
\, .
\eea
where we have omitted the vertex weight functions and the $\alpha,z$'s 
integrals (exactly the pre-factor in front of the large bracket in 
(\ref{ggkriz})). Integration  over the momentum $k$ (with the shift
$k+ (aP+bq)/2 \rightarrow k$ then yields 
\be  \label{sss}
-\frac{i}{2}\, \frac{\Gamma(7)}{\Gamma(3)}\,
\int \frac{d^4k}{(2\pi)^4}\, 
\frac{1+\frac{2k\cdot (2P+q)}{4M^2+Q^2}}
{\left[c-k^2 -k\cdot (aP+bq)-i\epsilon\right]^7}=
\frac{\Gamma(5)}{4(4\pi)^2}\,
\frac{1-\frac{4(Pa+qb).(2P+q)}{4M^2+Q^2}}
{\left[c+\frac{1}{4}(Pa+qb)^2\right]^5}
\ee
This relation can be further simplified using (\ref{impo}) in both the 
numerator and the denominator 
\bea  \label{iq}
I(Q^2)&=&\frac{\Gamma(4)}{(4\pi)^2}\,
\int\limits_{0}^{1}dt \int\limits_{0}^{1}dx
\int\limits_{0}^{1}dy \int\limits_{0}^{\frac{1+y}{2}}ds\, \,
\frac{ x (1-x) t^3 (1-t)^2 (1-a)}
{\left[\frac{Q^2}{4}b(b-a-1)-r \right]^5}
 \\
r&=&- (1-a^2)\frac{M^2}{4} + (1-t)m^2+ t(1-x) \alpha_{1}+ t x \alpha_2 \quad ,
\nn\\
a&=& t [z_1+x(z_2-z_1)] + (1-t)y \, ,
\nn \\
b&=& tx(1+z_2)t+ 2s(1-t) \, , \nn
\eea
It can be shown that for $Q^2\geq 0$ the denominator is nonzero. The 
function $I(Q^2)$ can be easily calculated numerically. Including the 
missing pre-factors the elastic electromagnetic form factor is given 
by: 
\begin{eqnarray}
G(Q^{2})&=&\int_{0}^{\infty}d\alpha_{1}d\alpha_{2}
\int_{-1}^{1}dz_{1}dz_{2}\, \rho^{[2]}(\alpha_1,z_1)\rho^{[2]}(\alpha_2,z_2)
\, I(Q^2) \quad .
\label{gqres}
\end{eqnarray}


%

\newpage

\begin{center}
\small{\begin{tabular}{|c|c|c|c|c|c|}
\hline \hline
$m_3/m  $&  $\eta=0$  & $\eta=0.2 $ & $\eta=0.5$ 
& $\eta=0.8 $ & $\eta=0.999$ \\
\hline \hline
0            & 1.9998 & 1.954  & 1.592 & 0.9067 & 0.03322 \\
\hline
0.5          & 2.5663 & 2.498  &  2.142 & 1.421 & 0.3873   \\
\hline
Ref.\cite{KUSAK3} &  2.5662& 2.4988 & -      & 1.4056& 0.3853   \\
\hline \hline
\end{tabular}}
\end{center}

\begin{center}
TABLE 1. Dimensioneless coupling $\tilde{\lambda}=g^2/(4\pi m)^2$ as a 
function of fraction of binding $\eta=\sqrt{P^2}/2m$ for two cases of 
exchanged mass $m_3$. The case $ m_3/m=0 $ is the 
Wick-Cutkosky model. The second case $m_3/m=0.5 $ is 
compared with the result obtained by Kusaka et al \cite{KUSAK3}. 
\end{center}

\

\

\
\begin{center}
\small{\begin{tabular}{|c|c|c|c|c|c|}
\hline
\hline
 $N_z*N_{\alpha}:$  &  16*16  & 32*32  & 64*64 & 96*96 & $\infty$ \\
\hline
\hline
$\eta=0.999;\, m_3=0.50 $ & 0.3782 & 0.3754 & 0.3794 & 0.3816 & 0.3874\\
\hline
$\eta=0.950;\, m_3=1.00 $ & 1.310 & 1.341 & 1.355 & 1.360 & 1.371\\
\hline
$\eta=0.950;\, m_3=0.50 $ & 0.752 & 0.761 & 0.777 & 0.783 & 0.804\\
\hline
$\eta=0.950;\, m_3=0.10 $ & 0.306 & 0.350 & 0.375 & 0.385 & 0.409\\
\hline
$\eta=0.000;\, m_3=1.00 $  & 3.143 & 3.273  & 3.342 & 3.366 & 3.416 \\
\hline
$\eta=0.000;\, m_3=0.50 $  & 2.207 & 2.343  & 2.445 & 2.483 & 2.566  \\
 
\hline
\hline
\end{tabular}}
\end{center}

\begin{center}
TABLE 2. The coupling $\tilde{\lambda}=g^2/(4\pi m)^2$ for 
bare ladder BSE as a function of the number of  mesh-points. 
\end{center}

\

\

\

\begin{center}
\small{\begin{tabular}{|c|c|c|c|c|c|c|c|}
\hline
\hline
$\eta $   & 0.0   & 0.4    & 0.5   & 0.5  & 0.5 & 0.95 & 0.95 \\
\hline
$m_3/m $  & 1.0   & 0.25   &  1.0  & 2.0  & 4.0 &  0.1 & 1.0\\
\hline $\lambda$  
          & 3.416 &  1.77 &  2.928 & 4.911 & 9.997 & 0.409 & 1.371 \\
\hline Ref.\cite{TJON}  
          & 3.419 &  na   &  2.940 &  na  & na  & 0.416 & 1.371  \\

\hline
\hline
\end{tabular}}
\end{center}

\begin{center}
TABLE 3. Dimensioneless coupling $\tilde{\lambda}=g^2/(4\pi m)^2$ for
several selections of $ m_3/m=$ and fraction of binding $\eta=\sqrt{P^2}/2m$,
compared to the results of ref. \cite{TJON}. 
 
\end{center}


\newpage

\begin{figure}[t]
\centerline{  \mbox{\psfig{figure=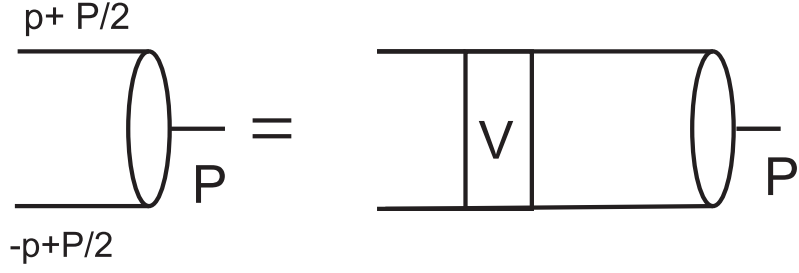,height=3.5truecm,angle=0}} }
\caption[99]{ \label{figBSE}Diagramatical representation of the BSE for
the bound state vertex function.}
\end{figure}

\begin{figure}[t]
\centerline{  \mbox{\psfig{figure=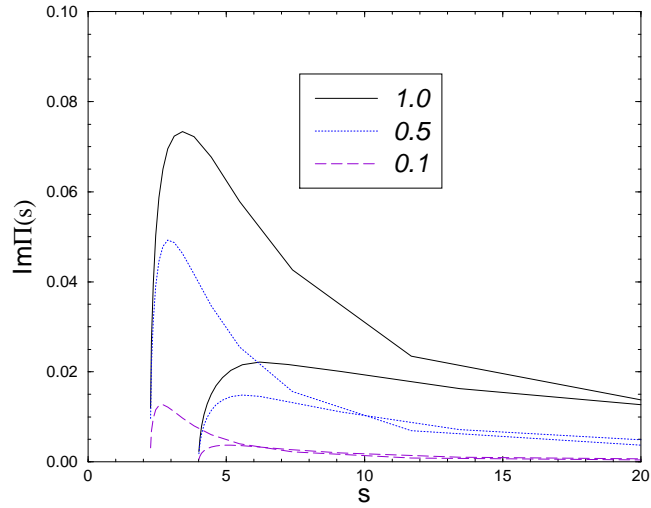,height=8.5truecm,angle=270}} }
\caption[99]{\label{figDSEsig} The imaginary part of the renormalized propagators 
for different values of coupling   
$\tilde{\lambda}$, calculated from the DSE in bare vertex approximation. Upper
curves are for particles $\Phi_{1,2}$ (which have identical self-energies,
lower ones for particle $\Phi_3$ .) }  
\end{figure}

\begin{figure}[t]
\centerline{  \mbox{\psfig{figure=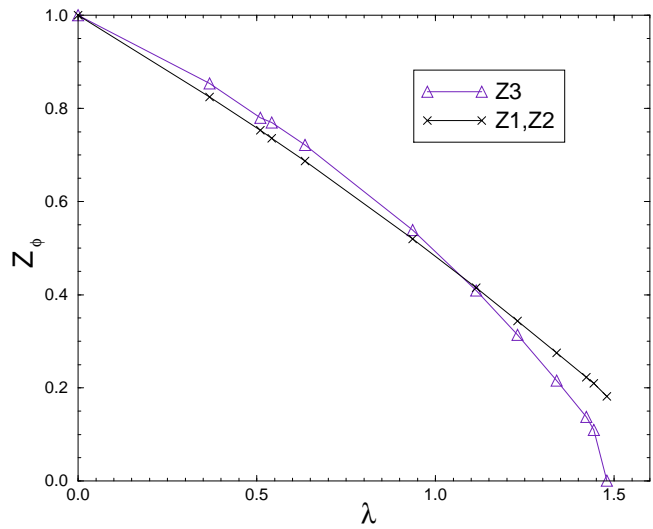,height=8.5truecm,angle=270}} }
\caption[99]{\label{figZs} The dependence of field strength renormalization constants 
on the coupling  $\tilde{\lambda}$.}  
\end{figure} 

\begin{figure}[t]
\centerline{  \mbox{\psfig{figure=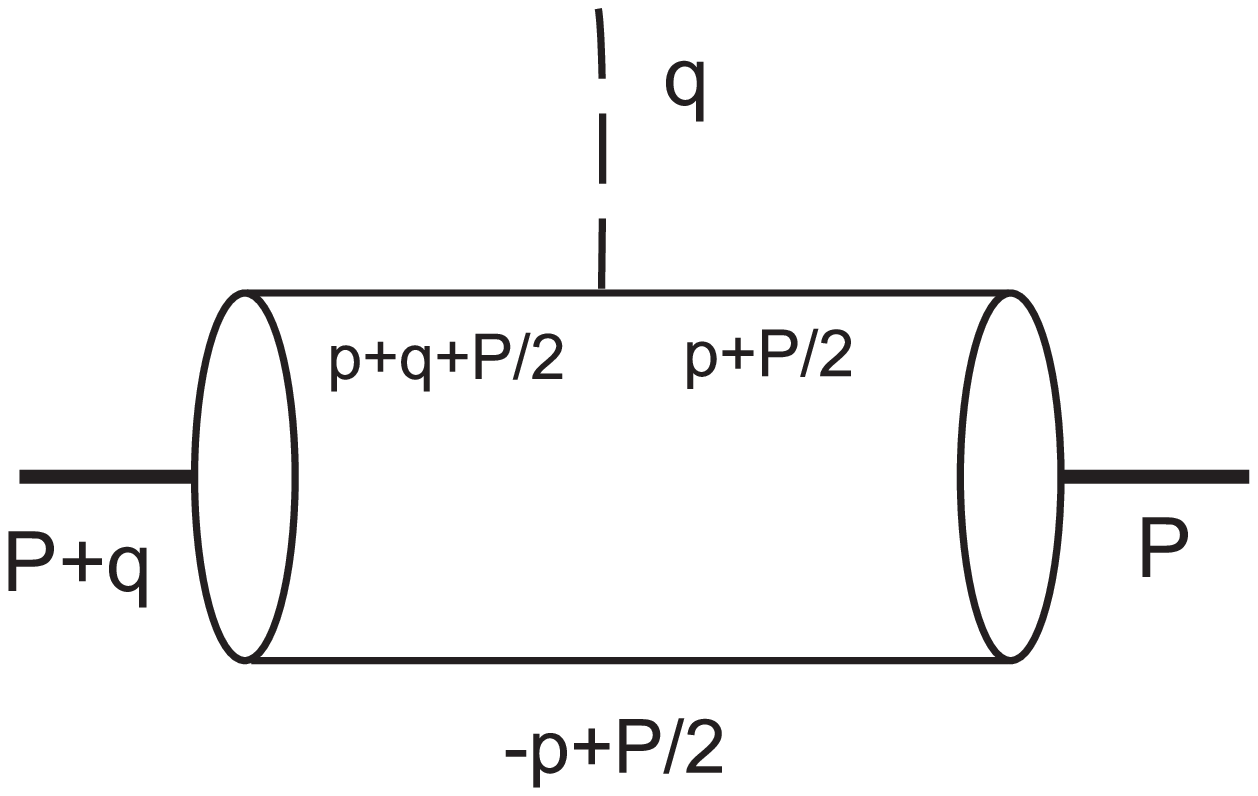,height=6.0truecm,angle=0}} }
\caption[99]{\label{figGmu} Diagrammatic representation of the e.m.\  current bound state
matrix element.}
\end{figure} 

\begin{figure}[t]
\centerline{  \mbox{\psfig{figure=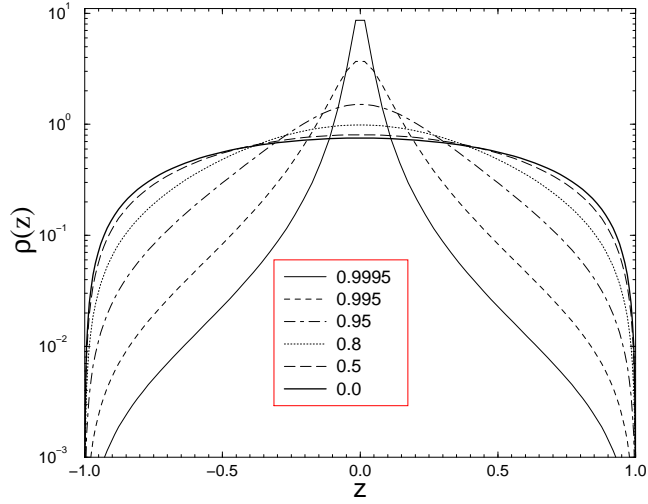,height=8.5truecm,angle=270}} }
\caption[99]{\label{figDSErho} The spectral function $\rho(z)$ of the bound-state vertex 
in the Wick-Cutkosky model for several values of $\eta=\sqrt{P^2}/2m$ .}
\end{figure}

\begin{figure}[t]
\centerline{  \mbox{\psfig{figure=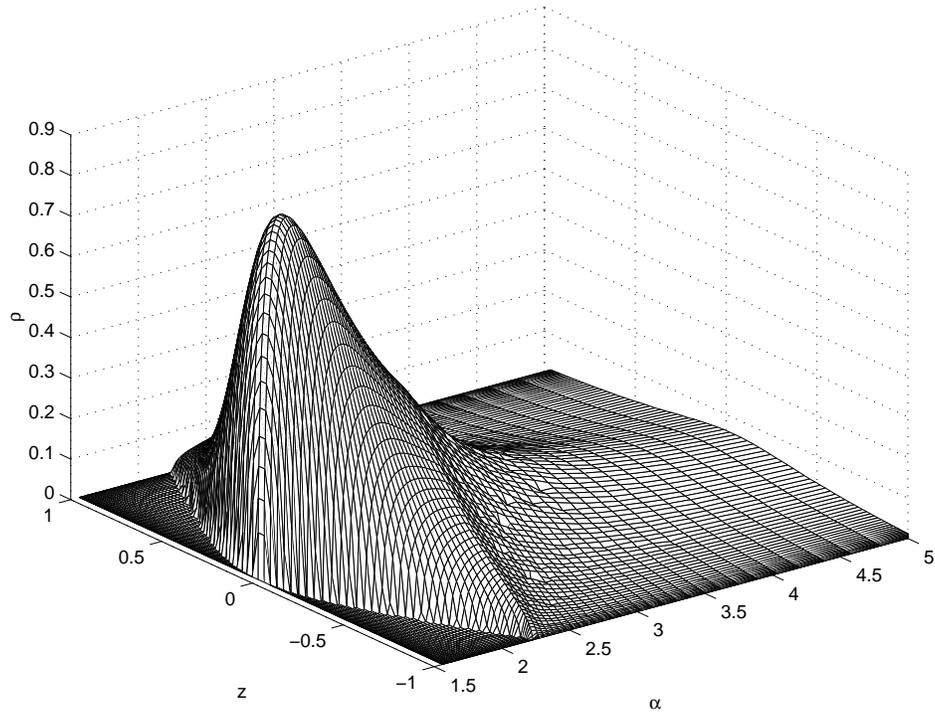,height=10.0truecm,angle=0}} }
\caption[99]{\label{figBSEbarelad1} The re-scaled weight function $\tilde\rho(\alpha,z)$ of the 
bound-state vertex for  $\eta=0.95$  calculated in bare ladder approximation. 
 The mass of the exchanged boson $m_3=0.5m$. }
\end{figure}

\begin{figure}[t]
\centerline{ \mbox{\psfig{figure=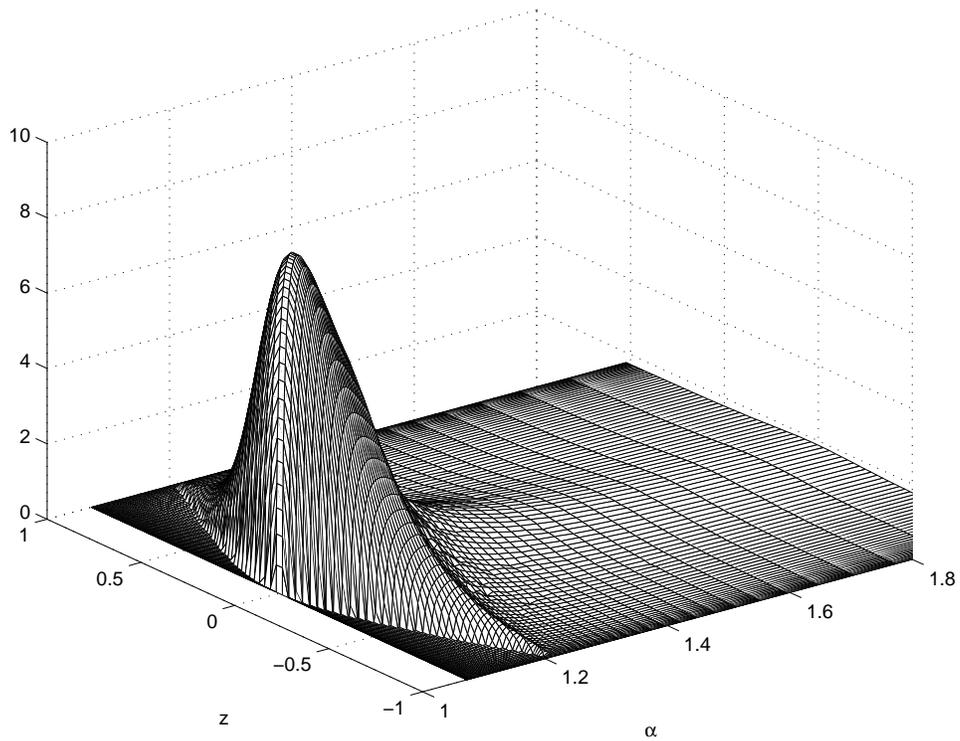,height=10.0truecm,angle=0}} }
\caption[99]{\label{figBSEbarelad2}  
The same as in previous figure, but for $m_3=0.1m$ and $\eta=0.95$.} 
\end{figure}

\begin{figure}[t]
\centerline{  \mbox{\psfig{figure=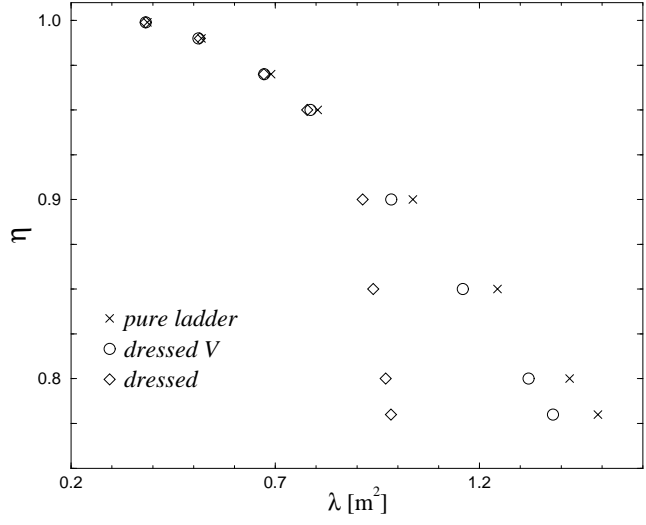,height=8.5truecm,angle=270}} }
\caption[99]{\label{figBSElam} The eigenvalues $\tilde{\lambda}$ calculated for the bare BSE, 
with dressed kernel $V$ and for dressed ladder BSE. Beyond the critical value
of coupling $\tilde{\lambda}_{crit}= 1.5$ only the bare solution is available. }  
\end{figure}

\begin{figure}[t]
\centerline{  \mbox{\psfig{figure=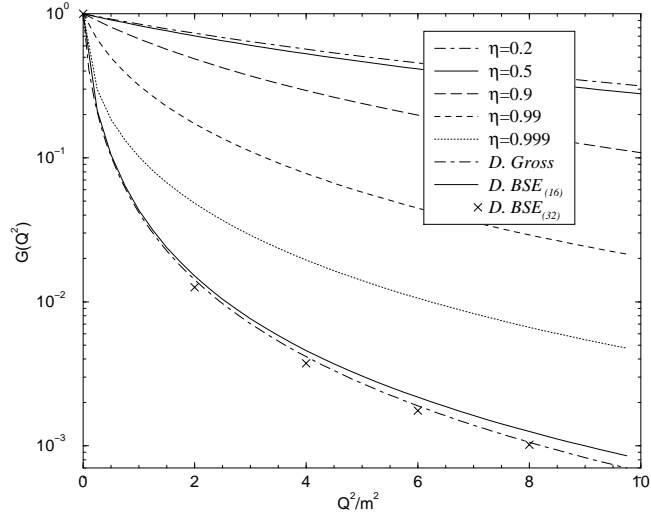,height=8.5truecm,angle=270}} }
\caption[99]{\label{figGfirst} The  behavior of the elastic electromagnetic form factors  for 
various bound states characterized by $\eta$.
The mass of exchanged particle is fixed to be $m_3=0.5m$, except for the scalar
deuteron case (D.), which is calculated for comparison using two different grids.}
\end{figure}

\begin{figure}[t]
\centerline{  \mbox{\psfig{figure=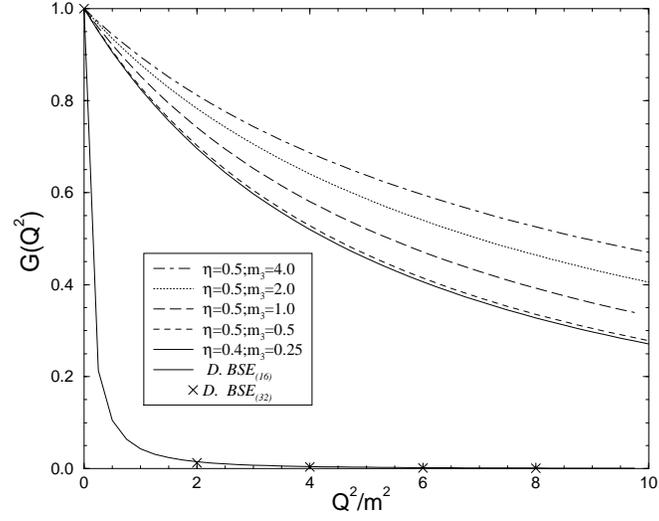,height=8.5truecm,angle=270}} }
\caption[99]{\label{figGsecond} Variation of the elastic electromagnetic form factors with the 
mass of the exchanged particle while $\eta=0.5$ is fixed.
The cases $\eta=0.4$, $m_3=0.25m $ and the scalar deuteron are included for comparison.}
\end{figure}

\end{document}